\documentclass[numberedappendix]{./emulateapj}
\newcommand{\mode}{emulateapj}
\usepackage{ifthen}
\newcommand{\mr}[1]{\mathrm{#1}}

\newcommand{\dd}[2]{\frac{d {#1}}{d {#2}}}

\newcommand{\tn}{\tablenotemark}

\newcommand{\m}{$^{-1}$}

\newcommand{\p}{^\prime}
\newcommand{\rhodm}{\rho_\mr{dm}}

\ifthenelse{\isundefined{\nonaas}}{}{

\bibpunct{(}{)}{;}{a}{}{,}

  \renewenvironment{thebibliography}[1]{%
    \begin{oldthebibliography}{#1}%
      \setlength{\parskip}{0ex}%
      \parindent 0ex%
      \setlength{\itemsep}{0ex}%
  }%
  {%
    \end{oldthebibliography}%
  }

\def\aj{{AJ}}			
\def\apj{{ApJ}}			
\def\apjl{{ApJ}}		
\def\apjs{{ApJS}}		
\def\ao{{Appl.~Opt.}}		
\def\aap{{A\&A}}		
\def\aaps{{A\&AS}}		
\def\mnras{{MNRAS}}		
\def\prd{{Phys.~Rev.~D}}	
\def\pasp{{PASP}}		
\def\nat{{Nature}}		

\def\physrep{{Phys.~Rep.}}   

}

\newcommand{\usemyrefs}{1}

\newcommand{\rdm}{r_\mr{d}}
\newcommand{\tablefontcommand}{\tabletypesize{\footnotesize}}
\newcommand{\fseall}{23345}
\newcommand{\fsedof}{23661}

\shortauthors{Mahdavi et al.}
\shorttitle{JACO}
\ifthenelse{\equal{\mode}{aastex}}
{
\newcommand{\previewspace}{}
\newcommand{\figfull}{6.7in}
\newcommand{\fighalf}{6.7in}
\newcommand{\figreallyhalf}{3.4in}
\newcommand{\figszwl}{5in}
\newcommand{\capfont}{\footnotesize}
\newcommand{\trot}{\rotate}
\newcommand{\dt}{deluxetable}
}{
\journalinfo{}
\submitted{Submitted May 16, 2006; Accepted March 2, 2007 for publication in \emph{The Astrophysical Journal}}
\newcommand{\previewspace}{\vspace{0.1in}}
\newcommand{\figfull}{7in}
\newcommand{\fighalf}{3.4in}
\newcommand{\figreallyhalf}{3.4in}
\newcommand{\figszwl}{\fighalf}
\newcommand{\dt}{deluxetable*}
\newcommand{\trot}{}
\newcommand{\capfont}{}
}

\begin{document}

\title{Joint Analysis of Cluster Observations: I. Mass Profile of Abell 478 from Combined X-ray, Sunyaev-Zel'dovich, and Weak Lensing Data}

\author{Andisheh Mahdavi, Henk Hoekstra and Arif Babul} \affil{Department of Physics
and Astronomy, University of Victoria, Victoria, BC V8W 3P6, Canada}
\author{Jonathan Sievers}
\affil{Canadian Institute for Theoretical Astrophysics, 60 St. George
St., Toronto, ON M5S 3H8, Canada}
\author{Steven T. Myers}
\affil{National Radio Astronomy Observatory, 1003 Lopezville Rd., Socorro, NM 87801}
\and
\author{J. Patrick Henry}
\affil{Institute for Astronomy, 2680 Woodlawn Drive, Honolulu HI 96822, U.~S.~A.}

\begin{abstract}
We provide a new framework for the joint analysis of cluster
observations (JACO) using simultaneous fits to X-ray,
Sunyaev-Zel'dovich (SZ), and weak lensing data. Our method fits the
mass models simultaneously to all data, provides explicit separation
of the gaseous, dark, and stellar components, and---for the first
time---allows joint constraints on all measurable physical
parameters. JACO includes additional improvements to previous X-ray
techniques, such as the treatment of the cluster termination shock and
explicit inclusion of the BCG's stellar mass profile. An application
of JACO to the rich galaxy cluster Abell 478 shows excellent agreement
among the X-ray, lensing, and SZ data. We find that Abell 478 is
consistent with a cuspy dark matter profile with inner slope
$n=1$. Accounting for the stellar mass profile of the BCG allows us to
rule out inner dark matter slopes $n>1.1$ at the 99\% confidence
level. At large radii, an $r^{-3}$ asymptotic slope is preferred over
an $r^{-4}$ behavior. All single power law dark matter models are
ruled out at greater than the 99\% confidence level.  JACO shows that
self-consistent modeling of multiwavelength data can provide powerful
constraints on the shape of the dark profile.
\end{abstract}


\section{Introduction}
\label{sec:introduction}
N-body simulations of our Cold Dark Matter (CDM) dominated universe
suggest that evolved galaxies and clusters of galaxies are contained
in dark halos with self-similar density profiles. In the collisionless
limit, $\rhodm(r/\rdm)/\delta = f(x)$. The characteristic density
$\delta$ and radius $\rdm$ depend on the halo mass and the
cosmological parameters, but the normalized profile $f(x)$ itself does
not
\nocite{NFW,Moore98,Ghigna00,Subramanian00,Fukushige01,Williams04,Merritt05}({Navarro}, {Frenk}, \& {White} 1997; {Moore} {et~al.} 1998; {Ghigna} {et~al.} 2000; {Subramanian}, {Cen}, \&  {Ostriker} 2000; {Fukushige} \& {Makino} 2001; {Williams}, {Babul}, \&  {Dalcanton} 2004; {Merritt} {et~al.} 2005).
The robustness with which collisionless N-body simulations predict a
universal profile makes the observational measurement of $f(x)$
particularly important. If we are to be satisfied that our
understanding of hierarchical structure formation is correct, we need
to use the best available data to test the theoretical dark matter
profiles in detail. 

Observations of $f(x)$ also have implications for self-interacting
dark matter (SIDM) as an alternative to the collisionless models. If the
weak interaction cross section of the particles in a halo is large
enough, substantial effects on the shape of the density profile should
appear within a Hubble time. Initially cuspy profiles should evolve
constant-density cores with a size $\sim 3\%$ of the virial radius
\nocite{Spergel00,Burkert00}({Spergel} \& {Steinhardt} 2000; {Burkert} 2000). It is an open question whether the SIDM
cores are stable and long-lived enough to be observable
\nocite{Balberg02,Ahn05}({Balberg}, {Shapiro}, \&  {Inagaki} 2002; {Ahn} \& {Shapiro} 2005) or whether the cores quickly collapse, forming
an $r^{-2}$ isothermal cusp \nocite{Kochanek00}({Kochanek} \& {White} 2000).  If the cores are
stable, measurements of $f(x)$ can provide constraints on the
interaction cross section and hence on the nature of dark matter
itself.

Rich, $\gtrsim 10^{14} M_\sun$ clusters of galaxies are excellent
laboratories for the measurement of $f(x)$ because of the large number
of independent techniques available for mass measurement. In this
series of papers, we examine the complex task of constraining $f(x)$
\emph{simultaneously} using the X-ray emitting intracluster medium
(ICM), the Sunyaev-Zel'dovich decrement, and weak gravitational
lensing maps. For the joint analysis of cluster observations (JACO),
we always begin with separate luminous and dark mass profiles, and
calculate the observables---the X-ray spectrum, the SZ decrement, or
the tangential lensing shear---directly from the underlying physical
model. In other words, our fits always take place in the data space,
rather than in mass, density or temperature space. This approach
allows for (1) deformation of the theoretical profiles to account for
the appropriate instrumental degradation at each wavelength and (2)
the ability to combine lensing, SZ, and X-ray data in a single grand
total fit.

Other works have examined the multiwavelength
approach. \nocite{MiraldaEscude95}{Miralda-Escude} \& {Babul} (1995) carried out the first comparison of
physical X-ray and strong lensing models for a sample of three rich
clusters; \nocite{Squires96b}{Squires} {et~al.} (1996) extended the comparison to weak
lensing. \nocite{Zaroubi98}{Zaroubi} {et~al.} (1998) and \nocite{Zaroubi01}{Zaroubi} {et~al.} (2001), using idealized X-ray
and SZ observations of simulated clusters, showed that the unique
axisymmetric deprojection of the gas density profile is possible by
combining the two types of data.  \nocite{Dore01}{Dor{\'e}} {et~al.} (2001), also using simulated
clusters, argued that the simultaneous deprojection of SZ and weak
lensing data is sensitive to the noise properties of the
data. \nocite{DeFilippis05}{De Filippis} {et~al.} (2005) and \nocite{Sereno06}{Sereno} {et~al.} (2006) use combined X-ray and
SZ data to place constraints on the triaxiality and inclination of the
intracluster medium assuming that they are single-component,
isothermal systems; they do not attempt to constrain the dark profile.
Assuming a \nocite{NFW}{Navarro} {et~al.} (1997) mass profile, \nocite{Laroque06}{LaRoque} {et~al.} (2006) use joint X-ray
and SZ data from 38 clusters to constrain the gas mass fraction.

JACO is the first method to use X-ray, SZ, and weak lensing data
simultaneously to constrain the shape of the dark matter profile in
clusters. This paper describes a test of this method on the cluster
Abell 478.  Chief among the assumptions allowing the extraction of the
dark matter profile is that the gas in hydrostatic equilibrium with
the overall gravitational potential. If hydrostatic equilibrium is
strongly violated, then determination of the structure of the dark
halo from the X-ray data alone becomes difficult.  Because ongoing or
recent cluster mergers are common, the identification of clusters that
are as close to prototypical (or ``boring'') as possible is valuable
for detailed mass measurement.  The difficulty of finding systems
suitable for hydrostatic analysis is evident in the latest Chandra and
XMM-Newton studies of relaxed clusters.

Clusters that appeared smooth and relaxed before Chandra and
XMM-Newton observations often contain features that make them unfit
for equilibrium analysis.  Examples include cold fronts
\nocite{Vikhlinin01,Bialek02,Dupke03,Hallman04}({Vikhlinin}, {Markevitch}, \&  {Murray} 2001; {Bialek}, {Evrard}, \& {Mohr} 2002; {Dupke} \& {White} 2003; {Hallman} \& {Markevitch} 2004) or shocks
\nocite{Markevitch01b,Markevitch02,Fabian03,Fujita04}({Markevitch} \& {Vikhlinin} 2001; {Markevitch} {et~al.} 2002; {Fabian} {et~al.} 2003; {Fujita} {et~al.} 2004) in recent
mergers. In clusters with X-ray substructure, the use of hydrostatic
equilibrium may yield incorrect results \nocite{Poole06}({Poole} {et~al.} 2006). In many
otherwise seemingly relaxed clusters, heating by a central AGN is
required to reconcile the short cooling times with the deficit of $k T
\lesssim 1$ keV gas in the quantities predicted by quasihydrostatic
cooling flow models
\nocite{Tamura01,Fabian01,Matsushita02,Peterson03,Kaastra04,Voit05}({Tamura} {et~al.} 2001; {Fabian} {et~al.} 2001; {Matsushita} {et~al.} 2002; {Peterson} {et~al.} 2003; {Kaastra} {et~al.} 2004; {Voit} \& {Donahue} 2005). Gas
in the region directly affected by the AGN is unlikely to be in
hydrostatic equilibrium, and the equilibrium equations may yield
incorrect results.

The most recent work shows, however, that many clusters can be
successfully modeled as equilibrium systems outside the region of
influence of the central AGN. In these regions, the total density
distribution frequently resembles a pure $n=1$ \nocite{NFW}{Navarro} {et~al.} (1997) profile, and
either a single or double $\beta$-model provides a successful
description of the X-ray surface brightness
\nocite{Schmidt01,Arabadjis02,Allen02,Lewis03,Buote04,Pointecouteau04,Buote05,Gavazzi05,Pointecouteau05,Vikhlinin06}({Schmidt}, {Allen}, \& {Fabian} 2001; {Arabadjis}, {Bautz}, \&  {Garmire} 2002; {Allen}, {Schmidt}, \& {Fabian} 2002; {Lewis}, {Buote}, \& {Stocke} 2003; {Buote} \& {Lewis} 2004; {Pointecouteau} {et~al.} 2004; {Buote}, {Humphrey}, \& {Stocke} 2005; {Gavazzi} 2005; {Pointecouteau}, {Arnaud}, \&  {Pratt} 2005; {Vikhlinin} {et~al.} 2006). However,
in a few cases, evidence for a significantly shallower $n \lesssim
0.5$ \nocite{Ettori02,Sanderson05,Voigt06}({Ettori} {et~al.} 2002; {Sanderson}, {Finoguenov}, \&  {Mohr} 2005; {Voigt} \& {Fabian} 2006) exists, echoing similar
results in several strong gravitational lensing studies
\nocite{Tyson98,Sand02,Sand04}({Tyson}, {Kochanski}, \&  {dell'Antonio} 1998; {Sand}, {Treu}, \& {Ellis} 2002; {Sand} {et~al.} 2004).

The substantial progress allowed by Chandra and XMM-Newton data leads
us to consider whether the constraints on the dark matter structure
could benefit from a refinement of the mass fitting technique and the
use of optical and radio data in a simultaneous fit.  In this paper we
show that through the use of a multiwavelength modeling technique
specifically aimed at constraining the the dark matter (as opposed to
the total) mass profile, 
%
enhanced
constraints on the slope of the dark mass distribution in clusters are
possible. In \S\ref{sec:technique} we motivate the technique and
describe it in detail. We apply the technique to the rich cluster
Abell 478 in \S\ref{sec:analysis}, and compare the results with
previous work and N-body models in \S\ref{sec:discussion}.  We
conclude in \S\ref{sec:conclusion}.

\section{Method}
\label{sec:technique}

\subsection{General Principles}  

The purpose of JACO is to constrain the shape of the dark matter
profile in clusters of galaxies via a single multi-parameter fit to
X-ray, lensing, and Sunyaev-Zel'dovich data. The broad features of our
technique are:

\emph{Separation of the mass model into the gaseous, stellar, and dark
components}. Rather than fitting for the total gravitating mass, JACO
splits the potential into three separately modeled components, thus
guaranteeing their positivity. As a result JACO is incapable of
generating unphysical mass profiles as is sometimes the case with
parameterized X-ray temperature profiles \nocite{Pizzolato03,Buote04}({Pizzolato} {et~al.} 2003; {Buote} \& {Lewis} 2004).
The stellar mass contributes a substantial part of the gravitating
mass within 3\% of the cluster virial radius \nocite{Sand04}({Sand} {et~al.} 2004), an
important but sometimes neglected effect when testing for the presence
of a SIDM constant density core.

\emph{Direct constraints from uncorrelated data.} We conduct minimal
data processing. We use the mass models to calculate, project, and
PSF-distort theoretical spectra, and then fit these to the measured
X-ray count spectra. We avoid deprojecting the data, thus guaranteeing
fits to uncorrelated data. The gas temperature is handled as an
internal variable, so that we skip the temperature fitting stage
altogether. For Sunyaev-Zel'dovich measurements, we calculate and fit
the uncorrelated Fourier modes of the decrement directly.

\emph{No X-ray temperature weighting}. Because the mass models are fit
directly to the spectra there is no need to calculate emission- or
otherwise-weighted temperatures. Hydrodynamic N-body simulations show
that emission-weighted temperatures do not accurately reflect the
measured spectroscopic temperature \nocite{Mazzotta04,Rasia05}({Mazzotta} {et~al.} 2004; {Rasia} {et~al.} 2005); recent
methods for getting around this problem require the calculation of
theoretical weights that depend on the X-ray calibration files
\nocite{Vikhlinin06b}({Vikhlinin} 2006). JACO calculates 2D spectra directly from the
mass models. Temperature and density variations within any given
annulus are reflected in the projected 2D X-ray spectra without any
emission weighting.

What is truly new here is (1) the direct calculation of X-ray spectra
from the gravitational potential, (2) the use of this potential to
model Sunyaev-Zel'dovich and weak lensing observations simultaneously,
and (3) the ability to conduct a full, covariant error analysis on all
the physical parameters.  There are also other important features that
have not been implemented in any prior method, e.g. the strict
calculation of the mean molecular weight from the metallicity profile
(\S\ref{sec:fundamentals}) and the formal modeling for the termination
shock as the cluster boundary (\S\ref{sec:boundary}). Some of the
other aspects of JACO have been discussed before. For example,
\nocite{Markevitch97}{Markevitch} \& {Vikhlinin} (1997) separate the gravitational potential into gaseous
and dark components when modeling Abell 2256. \nocite{Fukazawa06}{Fukazawa} {et~al.} (2006) and
\nocite{Humphrey06}{Humphrey} {et~al.} (2006) employ the stellar profile of the BCG in the X-ray
fit. And ``Smaug'' \nocite{Pizzolato03}({Pizzolato} {et~al.} 2003) was the first technique to
incorporate direct fitting of projected spectra. JACO improves on
Smaug by fitting a complete physical model, and avoiding the use of
emission-weighted temperatures and coarsely gridded cooling
functions. While JACO incorporates relevant ideas from its
predecessors, it was written completely independently, as the physical
motivation behind JACO made reuse of other code either slow or
impossible.

\subsection{Fundamental Equations}
\label{sec:fundamentals}
We first write the emissivity of the intracluster plasma,
\begin{eqnarray}
\label{eq:plasma}
\label{eq:emissivity}
\epsilon_\nu & = & n_e n_H \lambda_\nu(T,Z) = \left\langle \frac{n_e}{n_H}
 \right\rangle n_H^2 
\lambda_\nu(T,Z), 
\end{eqnarray}
where $T$ is the temperature, $n_e$ is the electron density, and $n_H$
is the hydrogen nucleus density. The cooling function
$\lambda_\nu(T,Z)$ is weighted over all metals with relative
abundances fixed on the \nocite{Grevesse98}{Grevesse} \& {Sauval} (1998) scale; in this system the
absolute abundance is $Z$.  For $\lambda_\nu$, JACO users can choose
between the APEC and the MEKAL plasma codes, or they can provide their
own. Note that in ionization equilibrium $\langle n_e/n_H \rangle$ is
a function of metallicity; JACO recalculates it each time it derives
an X-ray spectrum for a new value of $Z$. The same applies to the mean
molecular weight $\mu$.  We find that if we did not recalculate $\mu$
and $\langle n_e/n_H \rangle$ each time (i.e., held them fixed at some
fiducial $\bar{Z}$), we would be making up to a 5\% systematic error
in the output spectrum.

In the approximation that the cluster contains a relaxed, spherical,
and stationary ideal gas, 
\begin{equation}
\frac{1}{\rho_g} \dd{}{r} \left(\frac{\rho_ g k T}{\mu m_p} \right)= -
\frac{G (M_\mr{d} + M_\mr{g} + M_\mr{s}) }{r^2}.
\label{eq:hydro}
\end{equation}
where $m_p$ is the proton mass, $M_d$ is the dark mass, $M_g$ is the
gas mass, and $M_s$ is the stellar mass contained with a radius $r$.
Axisymmetric solutions will appear in future work.

The most general solution to equation (\ref{eq:hydro}) is
\begin{equation}
k T(r) = \frac{\mu m_p}{\rho_g} 
  \left( P_c - \int_{r_c}^r
\frac{G M \rho_g}{r^{\prime 2}} \, dr^\prime \right)
\label{eq:hydrosol}
\end{equation}
where $P_c$ is the pressure at an arbitrary radius $r = r_c$. The
emitted spectrum within any volume then becomes
\begin{equation}
\int n_e n_H \lambda_\nu \left[\frac{\mu m_p}{\rho_g} 
  \left( P_c - \int_{r_c}^r
\frac{G M \rho_g}{r^{\prime 2}} \, dr^\prime \right)
,Z \right] dV
\end{equation}
Thus the spectrum is expressed purely as a function of mass and
metallicity. No temperature fitting is required.

Once we know the pressure, the Sunyaev-Z'eldovich decrement follows
directly from an integral of the pressure along the line of sight
\nocite{Birkinshaw99}({Birkinshaw} 1999):
\begin{equation}
y_{SZ} = \int \frac{ P \sigma_T }{m_e c^2} d z
\end{equation}
where $\sigma_T$ is the electron scattering cross section, and $m_e$
is the electron mass. Similarly, the reduced weak gravitational
lensing shear in the tangential direction may be calculated directly
from the total mass model \nocite{MiraldaEscude91}({Miralda-Escude} 1991):

\begin{eqnarray}
\langle g_T \rangle(R) & = & \frac{\bar{\kappa}(<R) - \kappa(R)}{1-\kappa(R)}\\
\kappa(R) & = & \frac{\Sigma(R)}{\Sigma_\mr{crit}} = \frac{1}{\Sigma_\mr{crit}}
\int_{-\infty}^{\infty} (\rho_d +\rho_g +\rho_s) dz \\
\bar{\kappa}(<R) & = & \frac{2}{\Sigma_\mr{crit} R^2} \int_0^R R^\prime \Sigma(R^\prime) dR^\prime \\
\Sigma_\mr{crit} & = & \frac{c^2}{4 \pi G} \frac{1}{D_A \beta_\mathrm{WL}}
\end{eqnarray}

Here $g_T$ is the reduced tangential shear, $\kappa$ is the
convergence, $\Sigma(R)$ is the surface mass density, $D_A$ is the
angular diameter distance to the cluster, and $\beta_\mathrm{WL}$ is a
measure of the redshift distribution of the lensed sources; see
\nocite{Hoekstra98}{Hoekstra} {et~al.} (1998) (\S6), and \S\protect\ref{sec:lensing} below.

\subsection{Boundary conditions}
\label{sec:boundary}
The values of $P_c$ and $r_c$ are influenced by the choice of the
outer cluster boundary. In the idealized case that the cluster is
infinite, $P_c$ and $r_c$ are fixed by the physical requirements that
the temperature be everywhere finite and that the gas density decline
monotonically with $r$. Then as $r \rightarrow \infty$, the term in
parenthesis must vanish at least as fast as $\rho_g$; otherwise, the
temperature at large radii will be infinite. Thus
\begin{equation}
P_c = \int_{r_c}^\infty \frac{G M \rho_g}{r^{\prime 2}} \, d r^\prime
\end{equation}
and
\begin{equation}
k T(r) = \frac{\mu m_p}{\rho_g} \int_r^\infty \frac{G M
\rho_g}{r^{\prime 2}} \, d r^\prime .
\label{eq:infiniT}
\end{equation}

Of course, real relaxed clusters are not infinite; at some point,
accretion from the surrounding intergalactic medium, and its
associated shocks, become important \nocite{Ostriker05}({Ostriker}, {Bode}, \& {Babul} 2005). To simulate
this we truncate the cluster at its virial radius. $P_c$ is then the
``surface pressure'' or pressure of the gas at the outer
boundary. Assuming that the density of the intergalactic medium and
the ICM at the termination point are the same, a useful estimate of
this boundary pressure is
\begin{equation}
P_c = q \rho_g v_\mr{circ}^2 = q \rho_g \frac{G M_\mr{tot}}{r_\mr{vir}}
\end{equation}
Where $r_\mr{vir}$ is the virial radius (and the radius of
termination), $v_\mr{circ}$ is the circular velocity of the halo, and
$q$ is a constant of order unity. Both $q$ and $v_\mr{circ}$ are a
function of the cosmological parameters. For the prevailing
$\Lambda$CDM cosmology, $q \sim 1.1$ and $r_\mr{vir}= r_{100}$, the
radius at which the cluster density is 100 times the critical density
of the universe \nocite{Pierpaoli01}({Pierpaoli}, {Scott}, \&  {White} 2001).

\subsection{Model Radial Profiles}
\label{sec:models}
We now require parameterized profiles to insert into the right-hand
sides of equations (\ref{eq:emissivity}) and (\ref{eq:hydrosol}).
Because we are eliminating the temperature from the two equations, no
assumptions regarding the form of the temperature profile are
required. However, we still require parameterized gas, dark, and
stellar mass profiles, as well as a metallicity profile.

For the gas profile, we use a ``triple'' $\beta$-model \emph{gas
density} profile. Most relaxed clusters of
galaxies studied to date have surface brightness distributions that
are well-described by a double $\beta$-model
\nocite{Mohr99,Hicks02,Lewis03,Jia04,Johnstone05}({Mohr}, {Mathiesen}, \& {Evrard} 1999; {Hicks} {et~al.} 2002; {Lewis} {et~al.} 2003; {Jia} {et~al.} 2004; {Johnstone} {et~al.} 2005); however certain
cases a triple version is required to achieve a good fit
\nocite{Pointecouteau04,Vikhlinin06}({Pointecouteau} {et~al.} 2004; {Vikhlinin} {et~al.} 2006, e.g.).

Our version of the triple $\beta$-model is
\begin{equation}
\rho_g = \sum_{i=1}^{3} \rho_i (1+r^2/r_{x,i}^2)^{-3 \beta_{i}/2},
\label{eq:doublebeta}
\end{equation}
Note that in contrast to previous work, we define the triple $\beta$
model in density, not in surface brightness. The asymptotic behavior
of our triple $\beta$ model is the same as the behavior of a surface
brightness model with the same $\beta$ parameters; only the details of
the transitions among the various regimes are different.  While the
multiple $\beta$ model has several more fitting parameters than a
broken power law, it is demonstrably better in many cases,
e.g. \nocite{Sun04}{Sun} {et~al.} (2004).

For the dark matter distribution, JACO allows for a choice of profiles
from the N-body and dynamical literature. A fundamental prediction of
most collisionless CDM simulations is that $\rhodm(r)$ rises inwardly
as $r^{-1}$ or steeper to radii comparable within the resolution
limit. The resulting model has become known as the ``universal'' dark
matter profile:
\begin{equation}
\rho_U = \rho_0 (r/\rdm)^{-n} (1+r/\rdm)^{n-3},
\label{eq:universal}
\end{equation}
where $x \equiv r/r_\mr{d}$. Traditionally, $n = 1$ \nocite{NFW}({Navarro} {et~al.} 1997) to 1.5
\nocite{Moore98}({Moore} {et~al.} 1998); however, we explore all values $0 \le n \le 2$ to
test for softened as well as cuspy dark matter potentials. A similar
profile that closely describes the distribution of stars in elliptical
galaxies is the ``$\gamma$ profile'' \nocite{Dehnen93,Tremaine94}({Dehnen} 1993; {Tremaine} {et~al.} 1994):
\begin{equation}
\rho_\gamma = \rho_0 (r/\rdm)^{-n} (1+r/\rdm)^{n-4}.
\end{equation}
The widely used \nocite{Hernquist90}{Hernquist} (1990) profile is a special case of the
$\gamma$ profile with $n=1$. Unlike the universal profiles, the
$\gamma$ family of profiles have simple analytical properties and
finite total mass.  We also allow for a simple, single power law mass
distribution by taking the limit $\rdm \rightarrow \infty$ in equation
\ref{eq:universal}.
 
Finally, the stellar matter distribution in the BCG can be modeled
either as a single $\beta$-model or as a modified three-dimensional
\nocite{Sersic68}{S\'ersic} (1968) profile:
\begin{equation}
\rho_S = \rho_0 \exp{-(r/r_s)^{\alpha_s}}
\label{eq:sersic}
\end{equation}
Thus $\alpha_s = 1/n_s$, where $n_s$ is the S\'ersic index such that
$n=4$ gives a \nocite{deVaucouleurs53}{de Vaucouleurs} (1953) profile. The shape parameters of
the this profile are measured from high quality optical photometry,
while the normalization is allowed to vary freely in the fit.  The BCG
profile allows us to define a constant baryon fraction ($f_b$) dark
matter model---this is a fiducial model where the dark matter density
is proportional to the sum of the gas and the BCG stellar densities.

The $\gamma$ and S\'ersic profiles above have closed-form integrated
mass profiles.  For the others, the enclosed mass can be expressed as
a rapidly computable incomplete beta function (see Appendix
\ref{sec:incompletebeta}).

The radial dependence of the metallicity distribution is not as well
constrained by previous studies. We do know that relaxed clusters of
galaxies show ample evidence of negative radial metallicity gradients,
but the exact form of the decline is not clear. We choose the general
distribution
\begin{equation}
Z(r) = \frac{Z_0 r_Z + Z_1 r }{r_Z + r},
\end{equation}
i.e., the metallicity progresses smoothly from $Z_0$ at the center to
$Z_1$ at the outer regions, with a rate controlled by $r_Z$. This
functional form is consistent with the abundance profiles derived from
observations \nocite{Irwin01,DeGrandi01,Sun03a,Buote03b,dePlaa05}({Irwin} \& {Bregman} 2001; {De Grandi} \& {Molendi} 2001; {Sun} {et~al.} 2003; {Buote} {et~al.} 2003; {de Plaa} {et~al.} 2004).

Each radial profile has 3 or more free parameters. The grand total fit
to the cluster involves more than a dozen free parameters.  Table
\ref{tbl:params} is an annotated list of all such parameters. Not all
parameters need to be fit at the same time; any number can be linked
together or frozen at specific values. In most applications, the
cluster redshift and the parameters relating to the shape of the
stellar light profile will be fixed at specific values, because it is
difficult or impossible to estimate them from X-ray data alone.  In
the case of poor quality data one can restrict the cluster model to a
single $\beta$-model or to a constant metallicity.

\begin{\dt}{llccl}
\tablecaption{List of Parameters in the Cluster Model \label{tbl:params}}
\tablewidth{0in}
\tablefontcommand
\tablehead{\colhead{Parameter} & \colhead{Sherpa Name} & \colhead{Units} & \colhead{Can be fit?} & \colhead{Comments}}
\startdata
$\Delta$            & contrast   & \nodata  & N  & Overdensity for masses and concentrations \\
\nodata             & precision  & \nodata  & N  & Precision goal \\ 
$\rho_d$            & darkmodel  & \nodata  & N  & Integer specifying dark mass model \\
$\rho_*$            & starmodel  & \nodata  & N  & Integer specifying stellar mass model \\
z                   & redshift   & \nodata  & Y  & Cluster redshift\\
$M_g$               & Mg         & $10^{14} M_\sun$  & Y & Gas mass within $r_\Delta$ \\
$r_{x_1}$           & rx1        & Mpc               & Y & Core radius of first $\beta$-model \\
$r_{x_2}$           & rx2        & Mpc               & Y & Core radius of second $\beta$-model \\
$r_{x_3}$           & rx3        & Mpc               & Y & Core radius of third $\beta$-model \\
$\beta_1$           & b1         & \nodata           & Y & Slope of first $\beta$-model \\
$\beta_2$           & b2         & \nodata           & Y & Slope of second $\beta$-model \\
$\beta_3$           & b3         & \nodata           & Y & Slope of third $\beta$-model \\
$f_2$               & xfrac2     & \nodata           & Y & Normalization of second $\beta$-model relative to the 1st \\
$f_3$               & xfrac3     & \nodata           & Y & Normalization of third $\beta$-model relative to the 1st \\
$M_d$               & Md         & $10^{14} M_\sun$  & Y & Dark mass within $r_\Delta$ \\
$n$                 & ndark      & \nodata           & Y & Slope of dark matter density profile\\
$r_0$               & r0         & Mpc               & Y & Size of inner constant-density dark matter core \\
$r_d$/$c$           & rdm1       & \nodata           & Y & Dark matter concentration\tn{a} \\
$Z_0    $           & z0         & \nodata           & Y & Metallicity at $r=0$ \\
$Z_\infty$          & zinf       & \nodata           & Y & Metallicity at large radii\\
$M_\mr{BCG}$        & Mstar      & $10^{14} M_\sun$  & Y & Total stellar mass of BCG \\
$\alpha_s$          & alpha      & \nodata           & Y & Index of 3D stellar light profile, $\exp{-(r/rs)^{\alpha_s}}$\\
$r_s$               & rs         & Mpc               & Y & Scale radius of stellar profile\\
$r_\mr{shock}$      & rshock     & Mpc               & Y & Radius of termination shock \\
\enddata
\tablenotetext{a}{The user can decide to fit the concentration $c \equiv r_\Delta/r_d$,
or else to fit $r_d$ itself.}
\tablecomments{Additional parameters representing the residual
 soft X-ray background may also be fit to the data.}
\end{\dt}

\subsection{Projection}

Deprojected spectral or temperature data points are always
correlated. Use of the $\chi^2$ statistic to fit correlated data is
complex, and problematic if the covariance matrix of the deprojected
data is unknown. A straightforward alternative is to project the model
spectra on the sky, and fit them to the uncorrelated data.  For
projection we follow the ``Smaug'' method \nocite{Pizzolato03}({Pizzolato} {et~al.} 2003) in
detail. If the inner and outer edges of the $i$th annulus are at
projected distance $R_{i-1}$ and $R_{i}$ respectively, then the
observed flux is:
\begin{equation}
F_\nu(i) = \frac{1}{4 \pi D_L^2} \int_{R_{i-1}}^{R_{i}} 2 \pi R d R
\int_{R}^{r_\mr{max}} \frac{2 \epsilon_{\nu^\prime} r \, d
r}{\sqrt{r^2-R^2}},
\label{eq:flux}
\end{equation}
Where $\nu^\prime = \nu / (1+z)$, $D_L$ is the luminosity distance to
a source at redshift $z$, and $r_\mr{max}$ is the outer limit of the
cluster. By switching the order of integration, it is possible to
further simplify the integral \nocite{Pizzolato03}({Pizzolato} {et~al.} 2003):
\footnote{Note that \nocite{Pizzolato03}{Pizzolato} {et~al.} (2003) have an extra $4 \pi$ in front
of their equation (3) because they take a different definition of the
emissivity.}
\begin{equation}
F_\nu(i) = \frac{1}{D_L^2} \int_{R_{i-1}}^{r_\mr{max}} r\,
\epsilon_{\nu^\prime} K(r) d r
\label{eq:final}
\end{equation}
where
\begin{equation}
K(r) = \left\{ \begin{array}{ll}
\sqrt{r^2-R_{i-1}^2}  &\mr{if\ r\ <\ R_i} \\
\sqrt{r^2-R_{i-1}^2}-\sqrt{r^2-R_i^2} &\mr{if\ r\ >\ R_i} \end{array} \right.
\end{equation}

\subsection{PSF Distortion}
\label{sec:psf}

A crucial step in comparing the projected model to the data is to
account for PSF distortions. The JACO approach is to manipulate the
data as little as possible, so we apply the distortion to the model
spectra $F_\nu(i)$ via a convolution matrix:
\begin{equation}
S_\nu(i) = \sum_{j=1}^N \Pi_\nu(i,j) F_\nu(j)
\end{equation}
where $S_\nu(i)$ is the final model spectrum for comparison with the
data, and $\Pi_\nu(i,j)$ contains the energy-dependent contribution of
each annulus to itself and every other annulus.  Given any set of
annuli it is possible to calculate $\Pi_\nu(i,j)$ and thus convolve
the model with the PSF. 

In general, the PSF of Chandra and XMM-Newton is a function of energy
and position. JACO makes an allowance for this. However, the shape of
the PSF is not strictly circular or even ellipsoidal. We take no
account of this anisotropy, instead treating the PSF as azimuthally
symmetric at every point. In-flight XMM-Newton calibration studies
have found that this treatment of the PSF yields sufficiently accurate
encircled energy profiles except for very bright point sources
\nocite{XMMmosPSF,XMMpnPSF}({Ghizzardi} 2001a, 2001b). The model PSF is given by
\begin{equation}
p_\nu(\phi) \propto (1+\phi^2/\phi_0^2)^{\zeta} \\
\label{eq:psf}
\end{equation}
where $\phi$ is the angular distance in arcseconds between the source
center and the position being considered. The shape parameters
$\phi_0$ and $\zeta$ are functions of energy and position across the
detector, and are empirically fit as polynomials in photon energy $E =
h \nu$ and source position $\theta$:
\begin{eqnarray} 
\phi_0 = c_1 + c_2 E + c_3 \theta + c_4 E \theta \\
\zeta = c_5 + c_6 E + c_7 \theta + c_8 E \theta 
\end{eqnarray}
Here $\theta$ is measured in arcminutes and $E$ is measured in
keV. The XMM-Newton calibration team has derived
\nocite{XMMmosPSF,XMMpnPSF}({Ghizzardi} 2001a, 2001b) values of $c_{1-8}$ from archival
observational data. While extensive treatment of the Chandra PSF
exists, no functional fits to the azimuthally averaged Chandra PSF
have been published. For this reason, we undertake ray-tracing MARX
\nocite{Wise97}({Wise} 1997) simulations of bright point sources at various off-axis
angles and energies.

We find that equation (\ref{eq:psf}) provides an acceptable
description of the Chandra PSF as well. We measure $c_{1-8}$ in a
similar manner to \nocite{XMMmosPSF}{Ghizzardi} (2001a), except that we use ray-traced
rather than in-orbit data. The polynomial coefficients for all
instruments are shown in Table \ref{tbl:psf}.

\begin{deluxetable}{lrrrr}
\tablecaption{Energy Dependence of the PSF \label{tbl:psf}}
\tablefontcommand
\tablehead{\colhead{Coefficient} & \colhead{ACIS} & \colhead{MOS1}
&\colhead{MOS2} & \colhead{pn}}
\startdata
$c_1$ & 1.137 & 5.074 & 4.759 & 6.636\\
$c_2$ & -0.054 & -0.236 & -0.203 & -0.305 \\
$c_3$ & 0.076 & 0.002 & 0.014 & -0.175 \\
$c_4$ & 0.034 & -0.018 & -0.023 & -0.007 \\
$c_5$ & 6.119 & 1.472 & 1.411 & 1.525 \\
$c_6$ & -0.254 & -0.010 & -0.005 & -0.015 \\
$c_7$ & -0.652 & -0.001 & -0.001 & -0.012 \\
$c_8$ & 0.051 & -0.002 &  0.000 & -0.001 \\
\enddata
\end{deluxetable}
Using this functional form for the PSF we can calculate the PSF
convolution matrix $\Pi_\nu(i,j).$ The details of the calculation
are given in Appendix \ref{sec:psfintegral}.

\begin{figure*}
\begin{tabular}{cc}
\resizebox{3.8in}{!}{\includegraphics{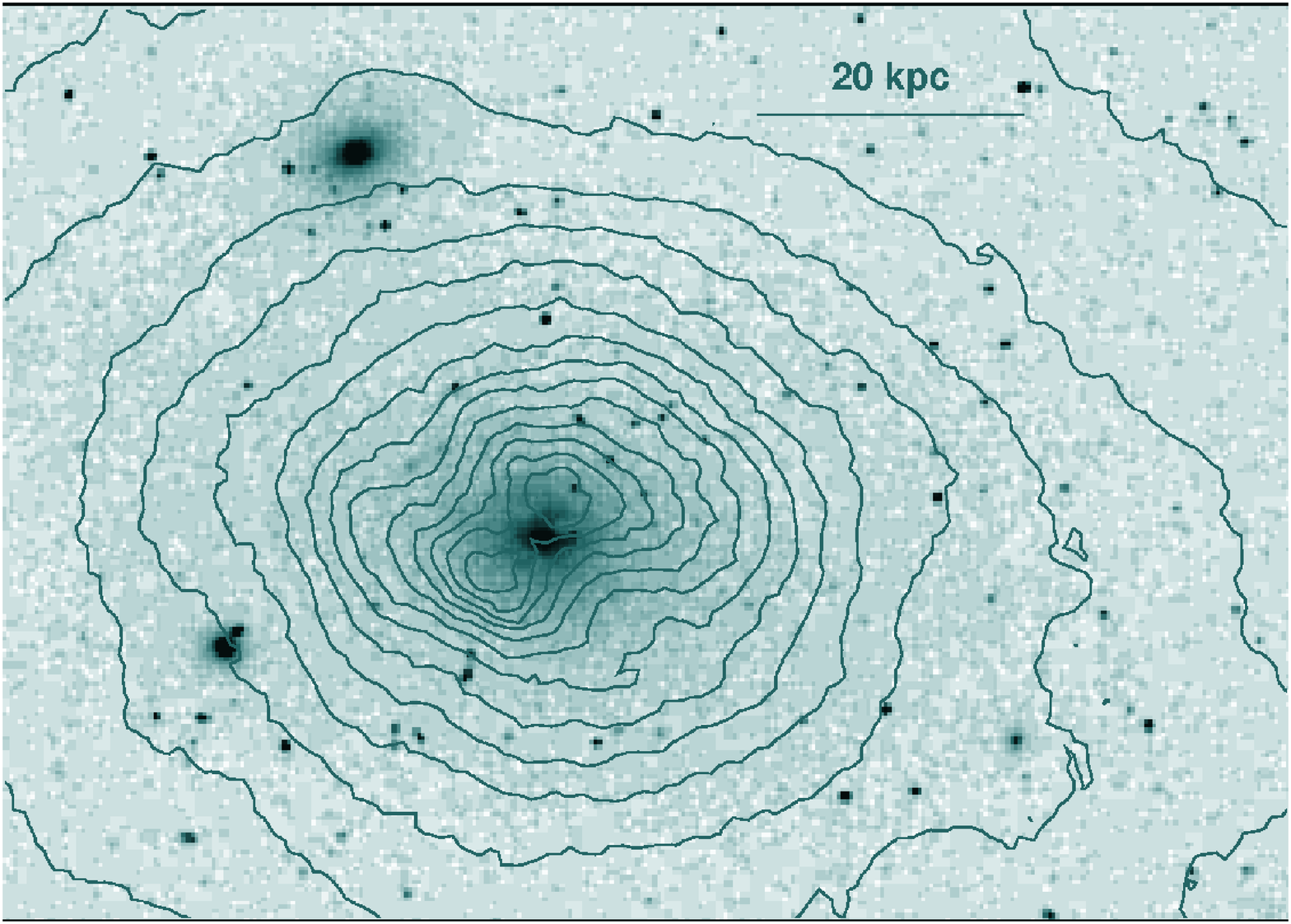}} &
\resizebox{2.8in}{!}{\includegraphics{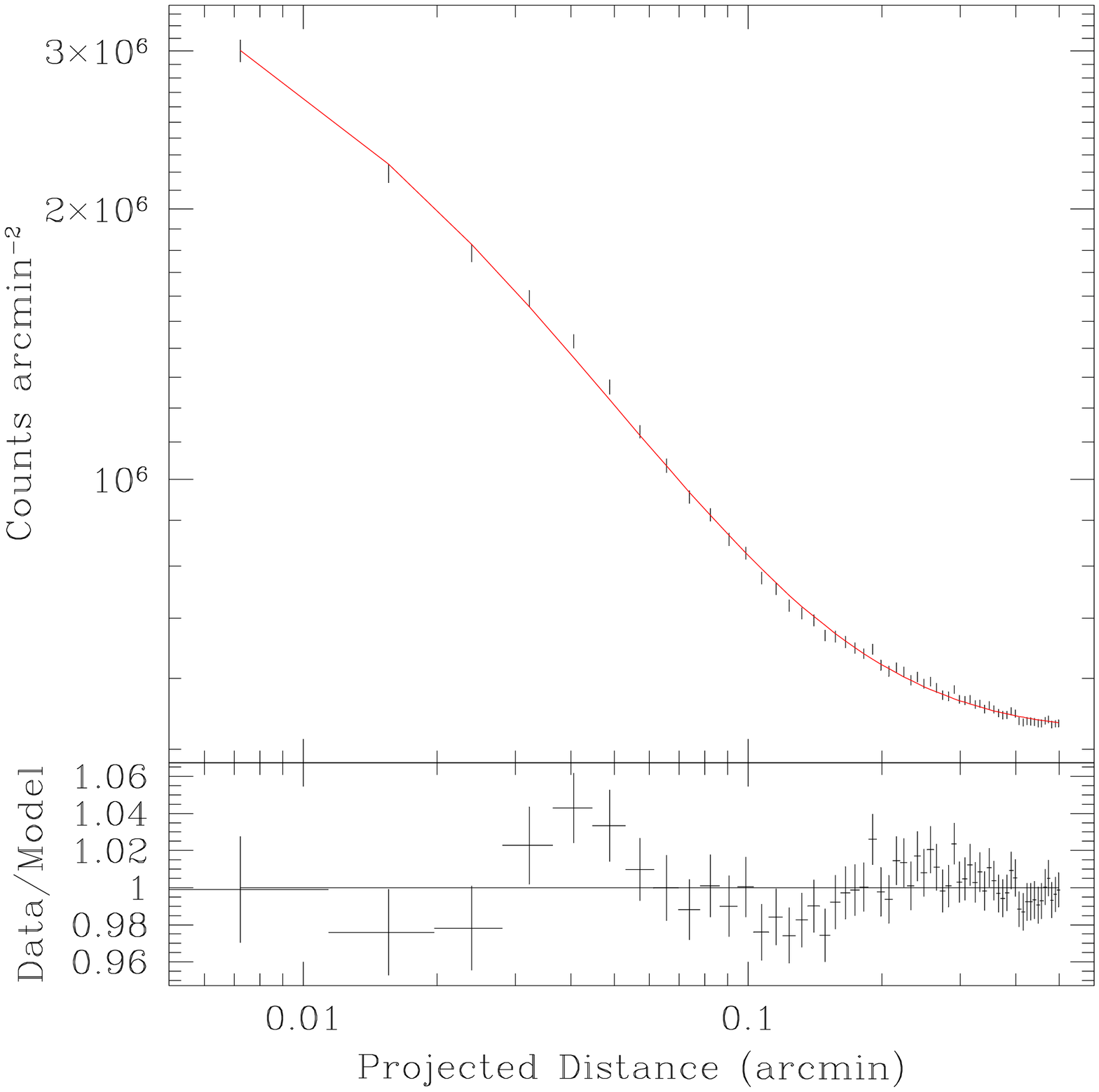}}
\end{tabular}
\caption{\capfont (\emph left) HST WFPC2 image of the core of Abell 478,
with superposed X-ray contours from the Chandra ACIS-S image. {\emph
right} Profile of the surface brightness plus background for the BCG,
as well as the best-fit S\'ersic model ($\alpha = 0.289$). \label{fig:hst}}
\end{figure*}

\subsection{Software Interface}

The JACO software package will soon be publicly available at
\url{http://jacocluster.sourceforge.net}. It consists of the following
components:
\begin{enumerate}

\item A core C language library which calculates the observed X-ray
spectrum, tangential shear, and Compton $y$ parameter from the input
dark, gas, and stellar mass distributions;

\item An interface which links the core library to the Sherpa data
analysis package \nocite{Sherpa1}({Freeman}, {Doe}, \&  {Siemiginowska} 2001) as a standard user-defined model, and
provides the necessary scripts for graphical viewing and fitting;

\item A parallelized interface, Hrothgar, which runs the core
JACO routine on multiprocessor facilities such as Beowulf
clusters. Hrothgar is a general purpose package and will soon be
available at \url{http://hrothgar.sourceforge.net}.

\item A set of optional data reduction scripts which process standard
archival X-ray data releases according to the procedures described in
Appendix \ref{sec:reduction}. Weak lensing and SZ data reduction is
up to the user. On request, the authors can also provide a routine for
transforming JACO Compton $y$ maps into interferometer observables.
The existing routines are customized for the Cosmic Background Imager
\nocite{Padin02}({Padin} {et~al.} 2002), but other arrays are easily accommodated.

\end{enumerate}


JACO offers a choice of numerical integration methods. The faster
method uses an adaptive Gauss-Legendre quadrature with 10 or 20
abscissae depending on the complexity of the integral. JACO also
offers adaptive Gauss-Kronod quadrature with convergence checking. The
latter method is slower but is guaranteed to converge to a given
accuracy. We find that the faster method produces model spectra with a
median accuracy of $0.1\%$.

\previewspace

\section{Application to Abell 478}

\label{sec:analysis}
\begin{figure*}
\begin{tabular}{cc}
\resizebox{\figreallyhalf}{!}{\includegraphics{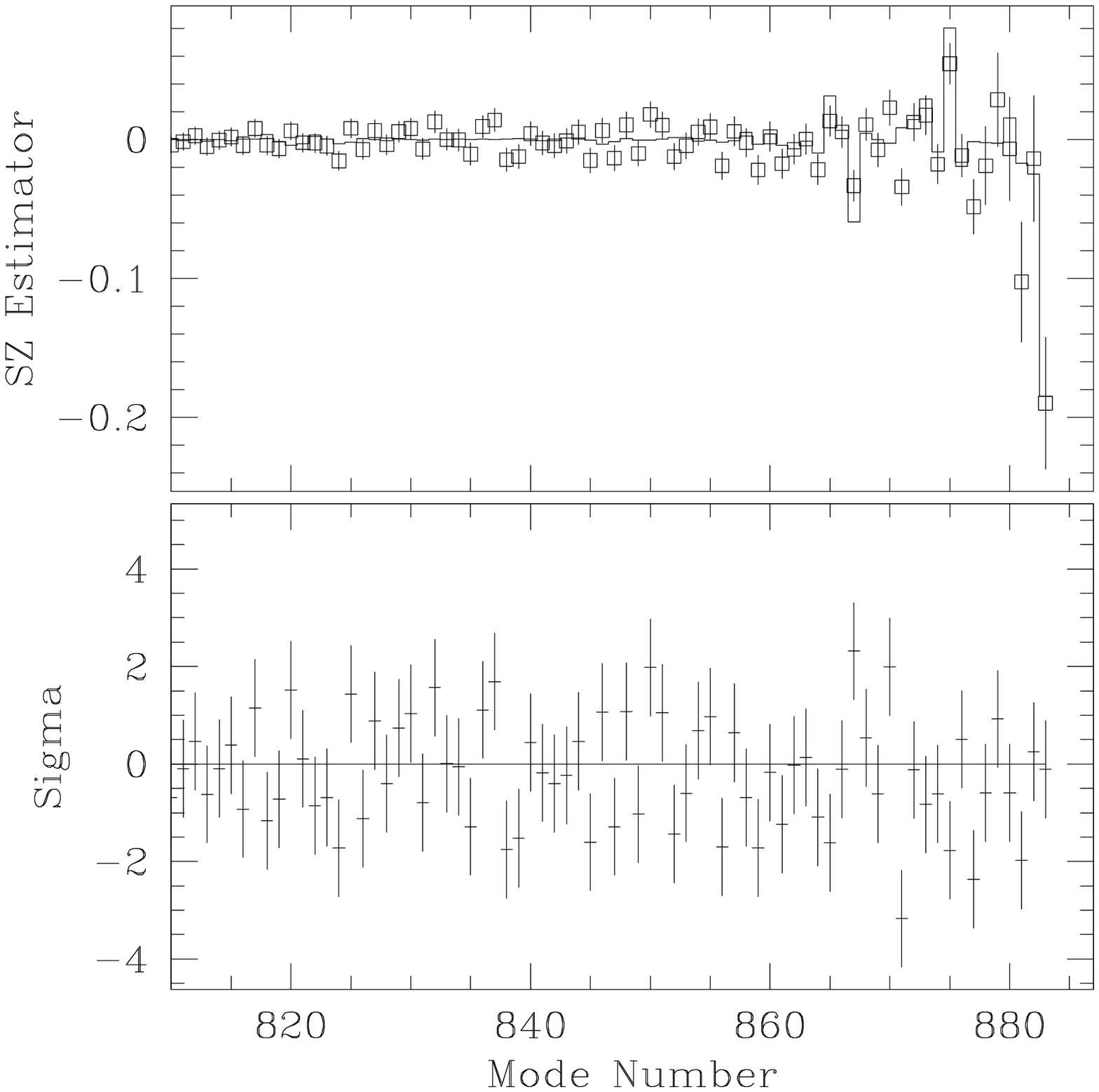}} &
\resizebox{\figreallyhalf}{!}{\includegraphics{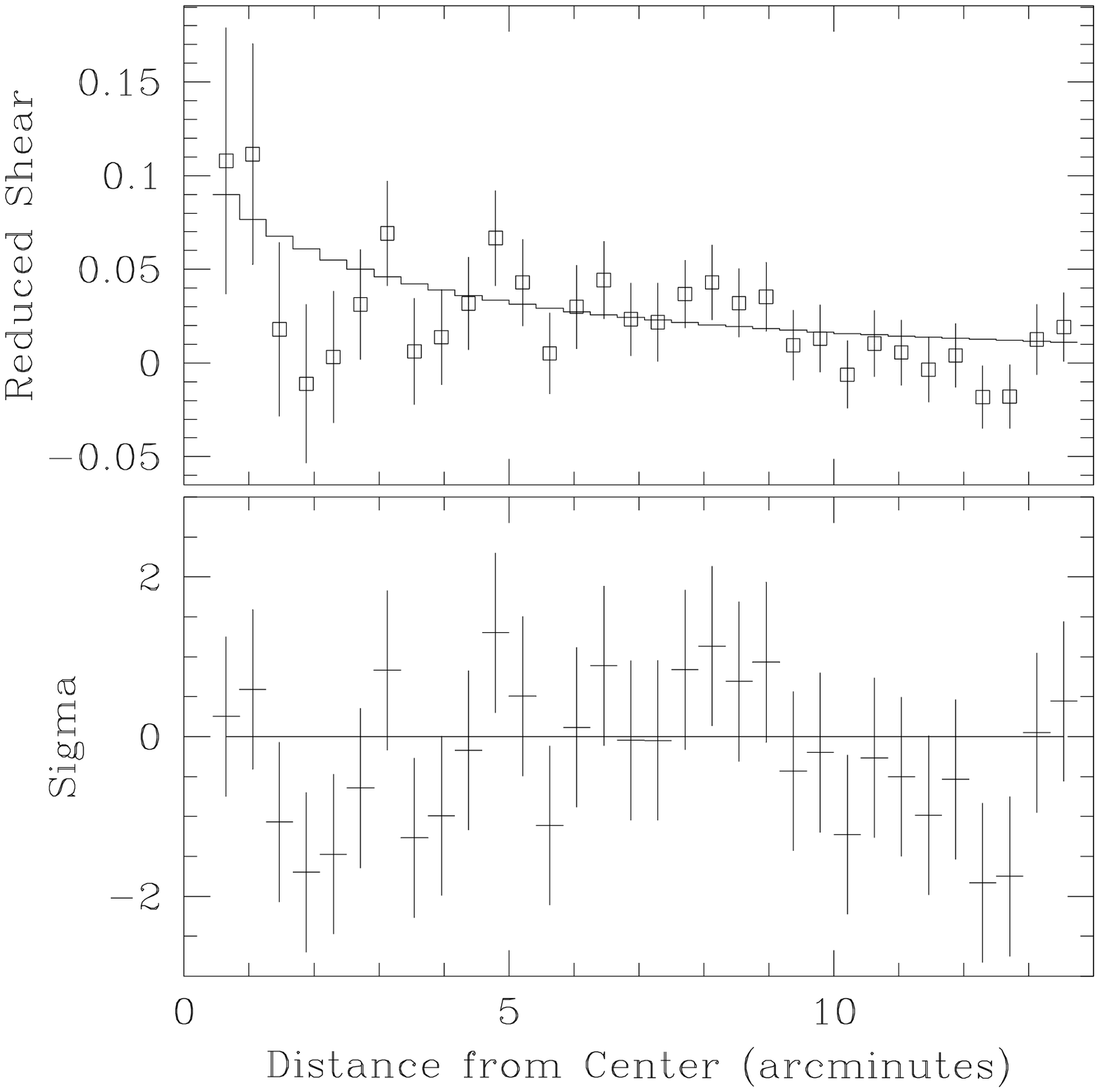}}
\end{tabular}
\figcaption{(\emph{left}) SZ data and residuals for
the best-fit JACO model; (\emph{right}) weak lensing data
and residuals for the best-fit JACO model \label{fig:wlszdata}}
\end{figure*}
\begin{figure*}
\resizebox{\figfull}{!}{\includegraphics{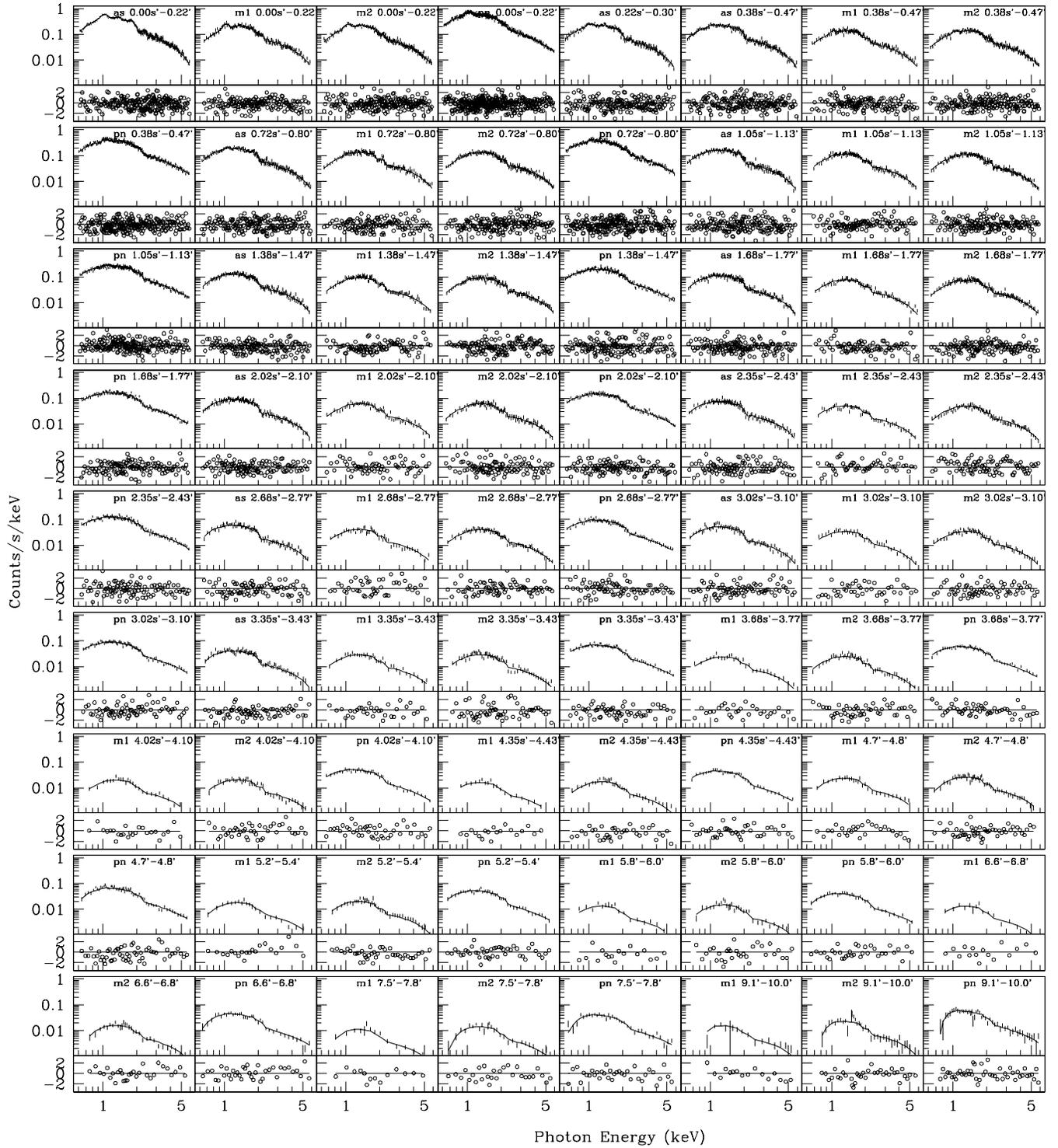}}
\figcaption{Spectra and best-fit models for Chandra ACIS-S (as),
XMM-Newton MOS1 (m1), MOS2 (m2), and pn cameras. Only 72 of the 268
total spectra are shown; the inner and outer radii of the extracted
annuli (in arcminutes) appear next to the instrument name.}
\label{fig:somespectra}.
\end{figure*}
\begin{figure*}
\resizebox{6.5in}{!}{\includegraphics{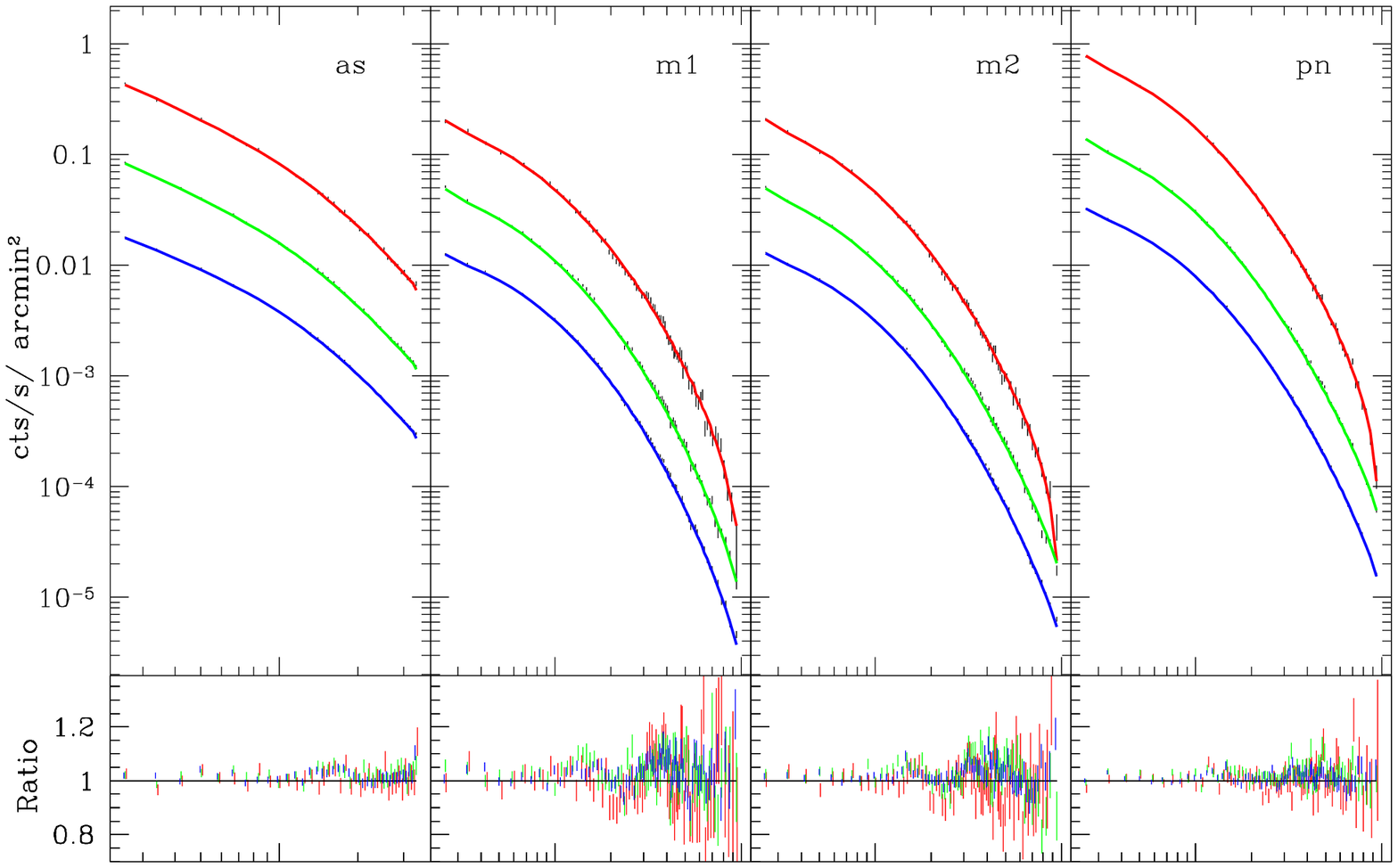}}
\resizebox{6.51in}{!}{\includegraphics{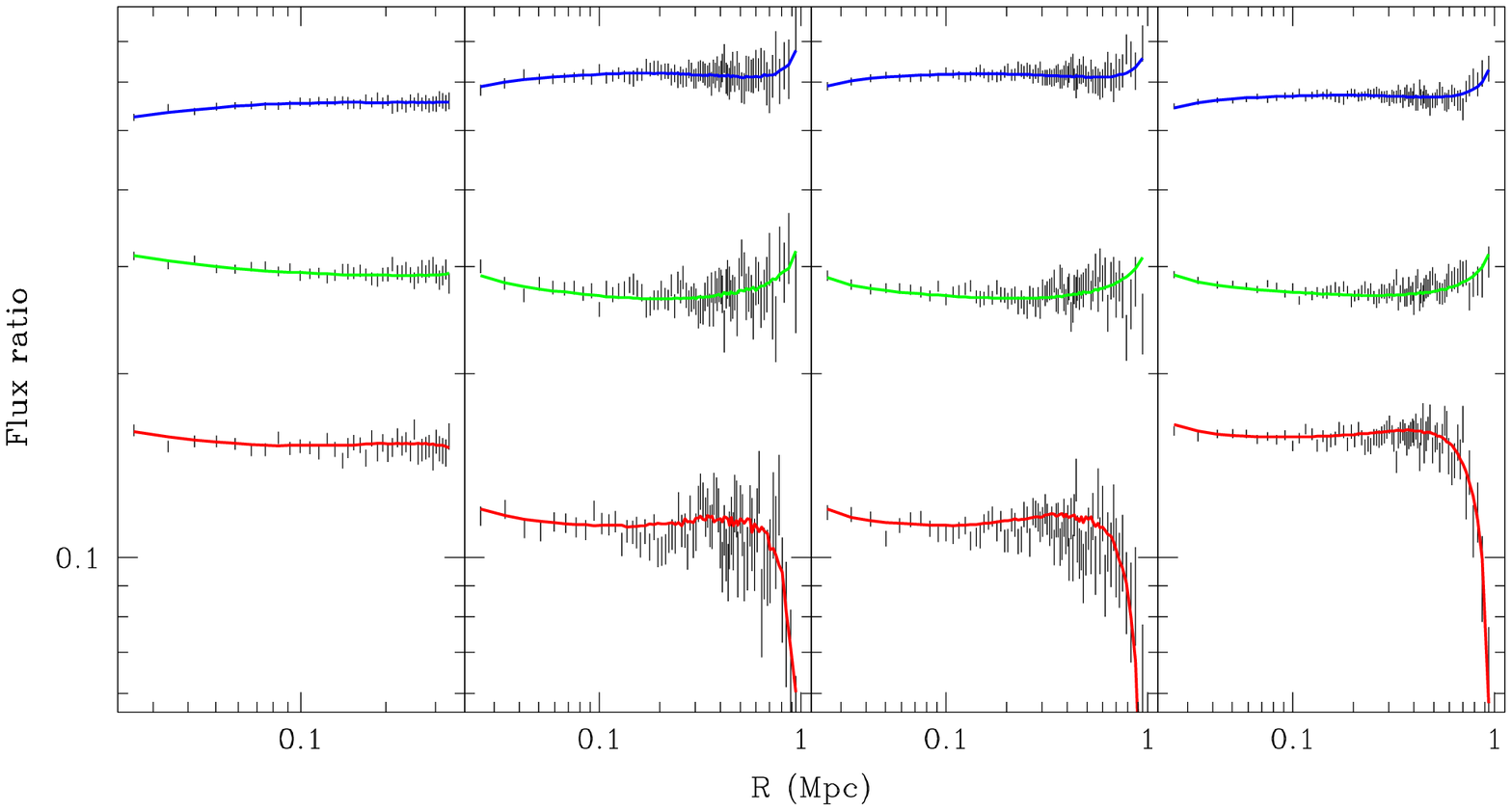}} \caption{\capfont Two
visualizations of the best-fit universal models (equation
\protect\ref{eq:universal}) for Abell 478. (\emph{top}) The spectra
are collapsed into three color regimes: 0.6-1.5 keV (upper line),
1.5-2.0 keV (middle line), and 2.0-6.0 keV (lower line). The count
rates are offset 0, -1, and -2 orders of magnitude, respectively, for
clarity. Data to model ratios are also shown. (\emph{bottom}) The
spectra and models are shown as color ratios. We show the ratio of
count rates in the three bands to the total count rate: 0.6-1.5 keV
(lower line), 1.5-2.0 keV (middle line), and 2.0-6.0 keV (upper
line). Each error bar here represents $\sim$ 50--70 of the fit
spectral data points, each of which in turn represents $\sim 50$
photons.
\label{fig:a478prof}}
\end{figure*}
Because the JACO technique is new, we apply it to a well-known, relaxed
cluster of galaxies.
Abell 478 is a rich cluster at z=0.088. The most recent available
studies with Chandra \nocite{Sanderson05,Voigt06}({Sanderson} {et~al.} 2005; {Voigt} \& {Fabian} 2006) and XMM-Newton
\nocite{Pointecouteau04}({Pointecouteau} {et~al.} 2004) suggest a peak ICM temperature of of $\sim
7-9$ keV, making this cluster one of the more massive known within $z
< 0.1$. For this analysis we assume $H_0 = 70$ km s\m\ Mpc\m, $\Omega_0
= 0.3$, and $\Omega_\Lambda=0.7$. With these values, $1\arcmin = 98.7$
kpc at the distance of Abell 478.

This cluster is an ideal test case for JACO, and highlights the
advantages of direct spectral fitting for hydrostatic mass
determination.  The Chandra and XMM-Newton observations record a total
of over $1,000,000$ source photons; yet \nocite{Pointecouteau04}{Pointecouteau} {et~al.} (2004),
\nocite{Sanderson05}{Sanderson} {et~al.} (2005), and \nocite{Voigt06}{Voigt} \& {Fabian} (2006) show that temperature profiles
of this cluster disagree when calculated with single temperature
emission-weighted plasma. We will show that the addition of optical
data (for the stellar mass profile and the weak gravitational lensing
shear) and radio data (for the Sunyaev-Zel'dovich decrement) improves
the constraints on the dark matter distribution in the cluster, and
bring the Chandra and XMM-Newton data into closer agreement.

\subsection{X-ray Data}

To model Abell 478, we fit JACO X-ray models to available XMM-Newton
and Chandra archival. There are four instruments in all: EPIC pn, EPIC
MOS 1 and 2, and Chandra ACIS-S. We used XMM-Newton observation
sequence 109880101 (126ks, of which 43ks were useful) and Chandra
ObsID 1669 (42ks, of which all was useful).  For the Chandra data, we
use CIAO 3.3 and CALDB 3.2.1. For the XMM-Newton data, we use the
latest CCF calibration release before August 2006.

We apply the generic reduction described in Appendix
\ref{sec:reduction}. We extract spectra in circular regions around the
X-ray centroid, which coincides in all datasets with RA(J2000) =
04:13:25.3, DEC(J2000) = +10:27:54. A total of 76 annuli are used, of
which Chandra covers only the innermost 40. The ACIS data are used to
choose the sizes of the innermost 40 annuli; we require at least 2500
background-subtracted counts per annulus. A similar requirement is
used to choose the sizes of the outer 36 annuli---2500
background-subtracted counts in the EPIC pn camera. The innermost
annulus is 13\arcsec\ (22 kpc) in radius; the annuli increase regular
by 5\arcsec\ (8 kpc) until a distance of 4.6\arcmin (450 kpc), at which
point they increase by 14\arcsec-40\arcsec\ until the final annulus at
9\arcmin\ (890 kpc).  The extracted spectra were fit over the 0.6-6 keV
energy range. The JACO plasma code was set to MEKAL.

The galactic absorption column varies with position on the sky
\nocite{Sun03a,Pointecouteau04,Vikhlinin05}({Sun} {et~al.} 2003; {Pointecouteau} {et~al.} 2004; {Vikhlinin} {et~al.} 2005). This effect is more
important for the XMM-Newton data, which cover a wider field than
Chandra; we take the gradient into account by allowing the hydrogen
column density to vary linearly with radius. Taking the variation of
$n_H$ with position into account still does not resolve the
temperature discrepancies between the Chandra and XMM-Newton
temperatures at intermediate radii
\nocite{Pointecouteau04,Vikhlinin05}({Pointecouteau} {et~al.} 2004; {Vikhlinin} {et~al.} 2005).

\subsection{Hubble Space Telescope Data}

Figure \ref{fig:hst} shows a 300s archival Hubble Space Telescope
WPFC2 image of the central $\sim 60$ kpc diameter region. The
corresponding HST observation number is U5A40901R, taken with the
$\sim 2000\AA$ bandpass F606W filter. Superposed are the Chandra ACIS
surface brightness contours.

The stellar mass of the BCG makes a non-negligible contribution to the
matter distribution within 50 kpc \nocite{Sand02,Sand04,Vikhlinin06}({Sand} {et~al.} 2002, 2004; {Vikhlinin} {et~al.} 2006),
and hence contributes to the equation of hydrostatic equilibrium at
all radii.  To model the stellar mass profile of the BCG, we fit its
HST surface brightness profile with a 3D \nocite{Sersic68}{S\'ersic} (1968) model
(equation \ref{eq:sersic}). Figure \ref{fig:hst} shows the best fit
model: $\alpha_\mr{BCG} = 0.289 \pm 0.002$ and $r_0 = 0.36 \pm 0.02
\arcsec$. This model is added to the gaseous and dark components when
conducting the grand total fit.

Using the photometric solution for the HST image, we calculate an
uncorrected F606W magnitude of $15.1\pm 0.1$ by integrating the
best-fit S\'ersic model to infinity. The galaxy is in a
high-extinction region; E(B-V) = 0.589 \nocite{Schlegel98}({Schlegel}, {Finkbeiner}, \&  {Davis} 1998). After
applying the extinction correction of $1.7$ mag and a k-correction
0.14 mag \nocite{Poggianti97}({Poggianti} 1997), we find that the total F606W luminosity
of the galaxy is $ 4.9 \pm 0.3 \times 10^{11}$ $L_\sun$.  This value
is important in constraining the average stellar mass-to-light ratio
of the BCG, $\Upsilon^*_\mr{BCG}$. For an evolved stellar population
with mean age $\gtrsim 3 Gyr$ in the wavelength regime of the F606W
filter, it is expected that $1 < \Upsilon^*_\mr{BCG} < 4 M_\odot /
L_\odot$ depending on the stellar initial mass function (IMF)
\nocite{Maraston98}({Maraston} 1998). The lower limit of 1 applies regardless of the
choice of IMF \nocite{Maraston98}({Maraston} 1998).

\subsection{Weak Lensing Data}
\label{sec:lensing}

The weak lensing measurements for A478 are based on archival data
taken with the CFH12k camera on the Canada-France-Hawaii-Telescope
(CFHT). The camera consists of an array of 6 by 2 CCDs, each 2048 by
4096 pixels. The pixel scale is $0.206\farcs$, which ensures good
sampling for the sub-arcsecond imaging data used here. The resulting
field of view is about 42 by 28 arcminutes, which is of great
importance when studying nearby clusters, such as A478.

The weak lensing analysis requires good image quality and therefore we
selected only data with seeing better than 0.8". This selection
yielded 9 exposures with 605s of integration each, resulting in a
total integration time of 5445s. Unfortunately only $R$-band
observations were available. This in principle complicates removal of
cluster members. However, our study of a large suite of $R$-band
observations of other clusters \nocite{Hoekstra07}({Hoekstra} 2007) has shown
that we can readily correct for this source of contamination using
the excess galaxy counts as a function of radius.

Detrended data (de-biased and flatfielded) are provided to the
community through CADC. 
We process this data through the analysis pipeline described in
\nocite{Hoekstra98}{Hoekstra} {et~al.} (1998), \nocite{Hoekstra00}{Hoekstra}, {Franx}, \&  {Kuijken} (2000), and \nocite{Hoekstra07}{Hoekstra} (2007). 
First we use the hierarchical peak
finding algorithm from \nocite{Kaiser95}{Kaiser}, {Squires}, \&  {Broadhurst} (1995) to find objects with a
significance $>5\sigma$ over the local sky. The peak finder gives
estimates for the object size, and we reject all objects smaller than
the size of the PSF. The remaining objects are analyzed, which yields
estimates for the size, apparent magnitude and shape parameters and
their measurement errors. The image is inspected by eye, in order to
remove areas that would lead to spurious detections. 

To measure the small, lensing induced distortions it is important to
accurately correct the shapes for PSF anisotropy, as well as for the
diluting effect of seeing. To characterize the spatial variation of
the PSF we select a sample of moderately bright stars from our images
\nocite{Hoekstra98}(e.g. {Hoekstra} {et~al.} 1998). Seeing circularizes the images, thus
lowering the raw lensing signal.  To correct for the seeing, we need
to rescale the polarizations to their ``pre-seeing'' value, as
outlined in \nocite{Hoekstra98}{Hoekstra} {et~al.} (1998). Our pipeline has been tested on
simulated images in great detail \nocite{Hoekstra98,Heymans06}(e.g. {Hoekstra} {et~al.} 1998; {Heymans} {et~al.} 2006).
These results suggest that we can recover the shear with an accuracy
of $\sim 2\%$ \nocite{Heymans06}({Heymans} {et~al.} 2006).

The catalog of objects with corrected shapes is used for the weak
lensing analysis. In the analysis presented here, we consider the
tangential distortion as a function of radius from the cluster
center. The resulting measurements, using galaxies with apparent
magnitudes $22<R<24.5$ are shown in Figure \ref{fig:wlszdata}.

The interpretation of the lensing signal requires knowledge of the
redshifts of the source galaxies. Based on our data alone, we do not
know the redshifts of the individual sources. The observed lensing
signal, however, is an ensemble average of many different galaxies,
each with their own redshift. As a result, it is sufficient to know
the source redshift distribution to compute the average value
$\beta=\langle D_{ls}/D_s\rangle$. Based on the Hubble Deep Fields, we
obtain $\beta=0.8$. We note that, because A478 is a low redshift
cluster, the inferred lensing mass is rather insensitive to the
detailed source redshift distribution as effectively all background
galaxies are far away.

\subsection{Sunyaev-Zel'dovich Data}

The SZ data were taken with the Cosmic Background Imager
\nocite{Padin02}(CBI; {Padin} {et~al.} 2002) over 11 nights as part of a complete sample of X-ray
luminous, low-redshift ($z<0.1$) clusters.  A478 is a remarkably clean
cluster, with no apparent point sources at 30 GHz in the CBI data.  By
far the dominant source of noise in the data is the CMB, which reduces
the overall significance of detection from 24.8 sigma, if only thermal
noise is considered, to 8.3 sigma.  See \nocite{Udomprasert04}{Udomprasert} {et~al.} (2004) for a
more complete treatment of observing the SZ effect with the CBI, as
well as a more detailed description of the A478 data used here.

To calculate $\chi^2$ from the SZ data, we first run the visibilities
through CBIGRIDR, the CBI CMB analysis pipeline \nocite{Myers03}({Myers} {et~al.} 2003).  This
compresses them into ~2000 "gridded estimators", and calculates the
noise and CMB covariances between the estimators, as well as source
vectors for source projection if so desired \nocite{Bond98}(e.g. {Bond}, {Jaffe}, \& {Knox} 1998).
This is done once per cluster, before fitting cluster models.  During
the fit, Compton $y$ maps for models are run through the CBI
simulation pipeline to generate predicted visibilities, which are then
run through CBIGRIDR to generate the predicted estimators.  Let
$\Delta$ be the estimators, $\Delta_M$ be the predicted estimators for
the current model, $S_{CMB}$ be the CMB covariance matrix, and $N$ be
the noise covariance matrix, then $\chi^2=(\Delta-\Delta_M)^T
(N+S_{CMB})^{-1} (\Delta-\Delta_M)$.  Since the noise and the CMB
covariance are independent of the cluster model, the inverse need be
taken only once.  In practice, we rotate into a space in which
$N+S_{CMB}$ is diagonal using $V$, the eigenvectors of $N+S_{CMB}$.
Let $\Delta^* \equiv V^T\Delta$, then $\chi^2$ reduces to $\sum
(\Delta^*_i-\Delta^*_{M,i})^2/\sigma_i^2$ where $\sigma_i^2$ is the
corresponding eigenvalue of $N+S_{CMB}$.  This lets us treat the
rotated estimators as independent, uncorrelated measurements of the
sky.  In practice, it takes a few minutes to calculate the eigenvector
rotation, and each $\chi^2$ evaluation (including running the
visibility pipeline and CBIGRIDR for estimators) takes $\sim 0.5$
seconds. The data for the mode numbers which contain most of the
cluster signal appear in Figure \ref{fig:wlszdata}.  We do not fit
mode numbers $< 600$, which are guaranteed not to contain any cluster
signal.

\subsection{Results}
\label{sec:results}

First we fit the JACO models separately to the Chandra and XMM-Newton
data.  We thereby gain insight into how the differences among the four
separate instruments affect the estimated physical parameters.  
\begin{figure*}
\begin{tabular}{cc}
\resizebox{\figreallyhalf}{!}{\includegraphics{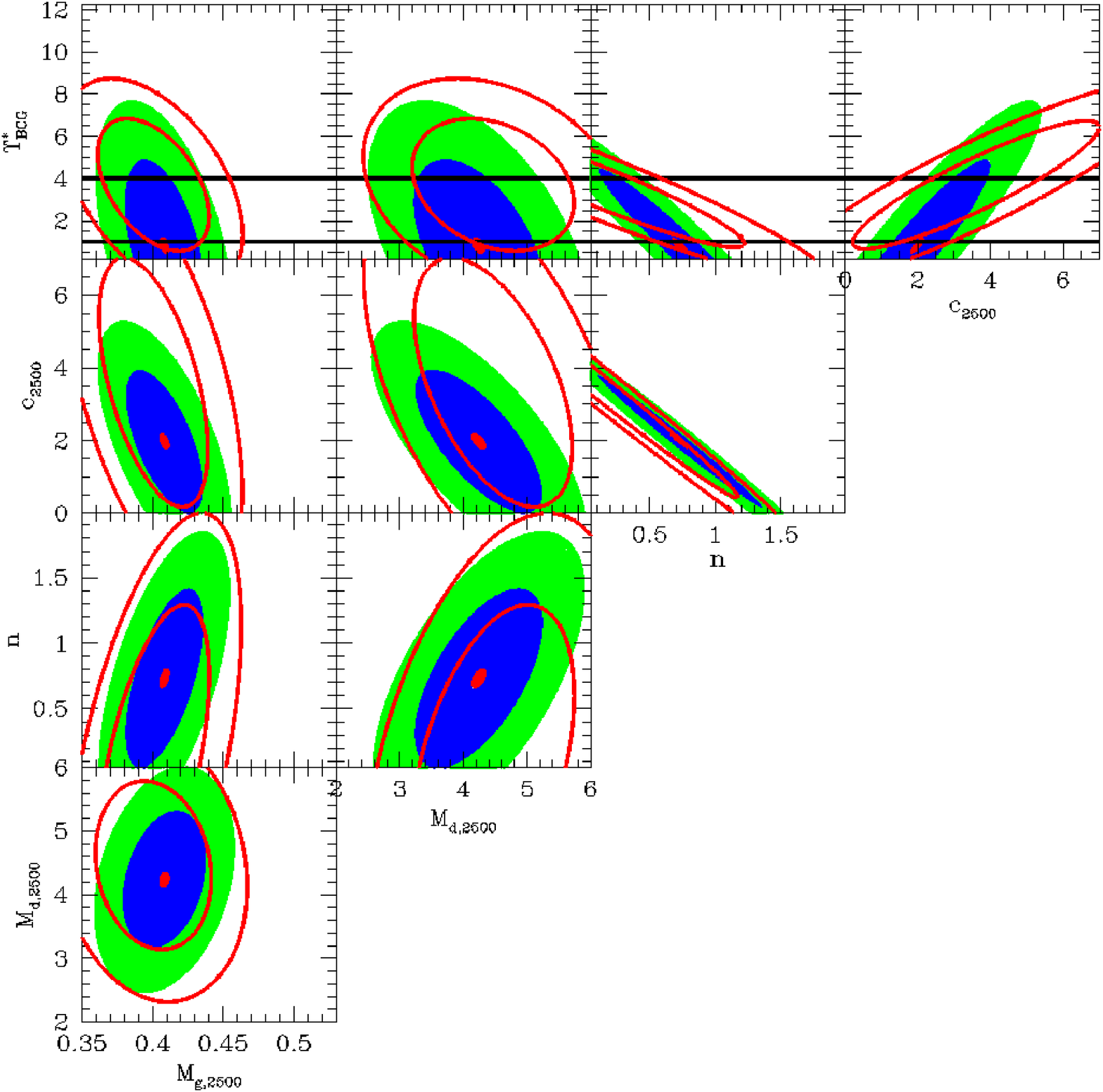}} &
\resizebox{\figreallyhalf}{!}{\includegraphics{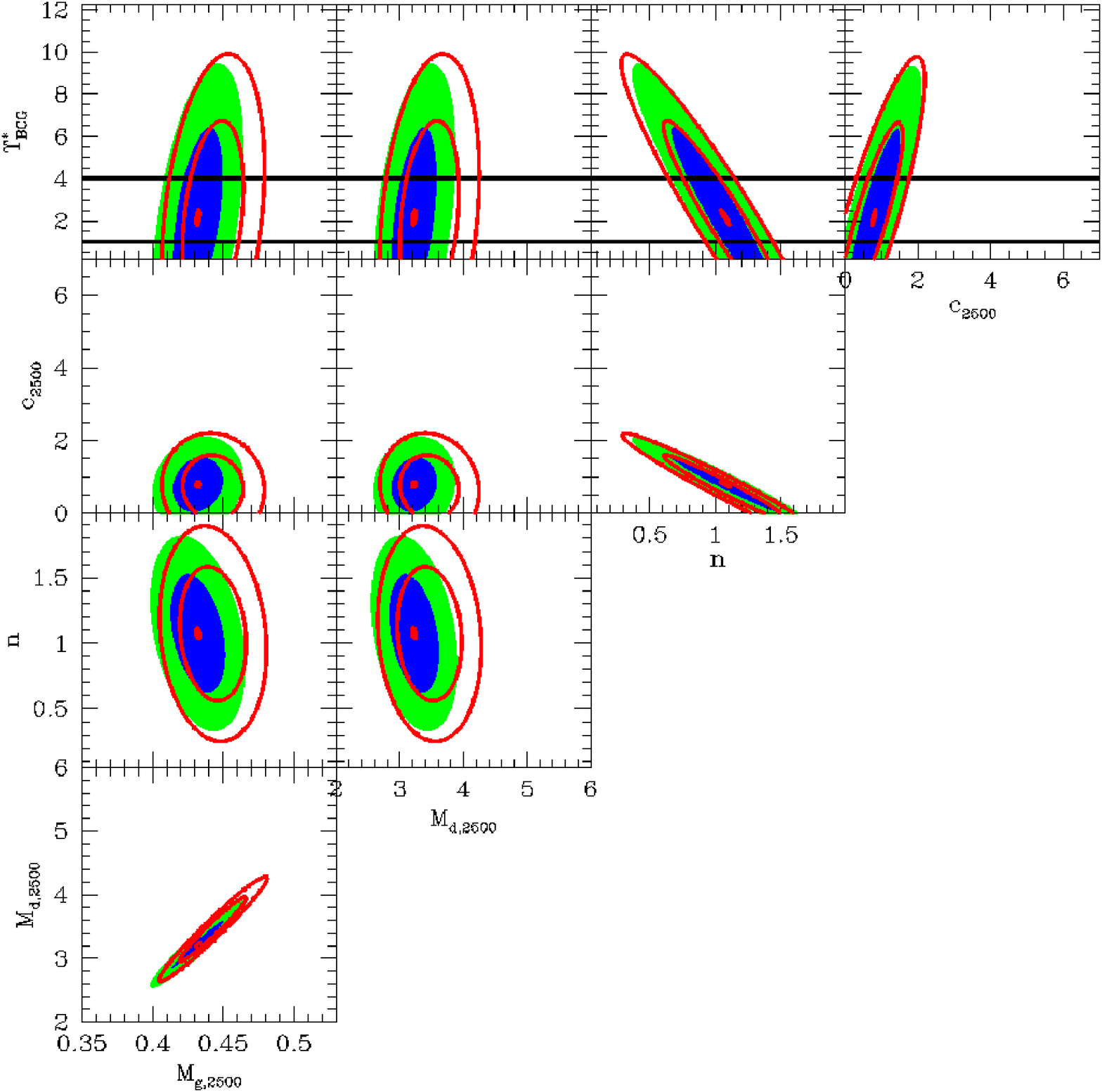}} \\
\resizebox{\figreallyhalf}{!}{\includegraphics{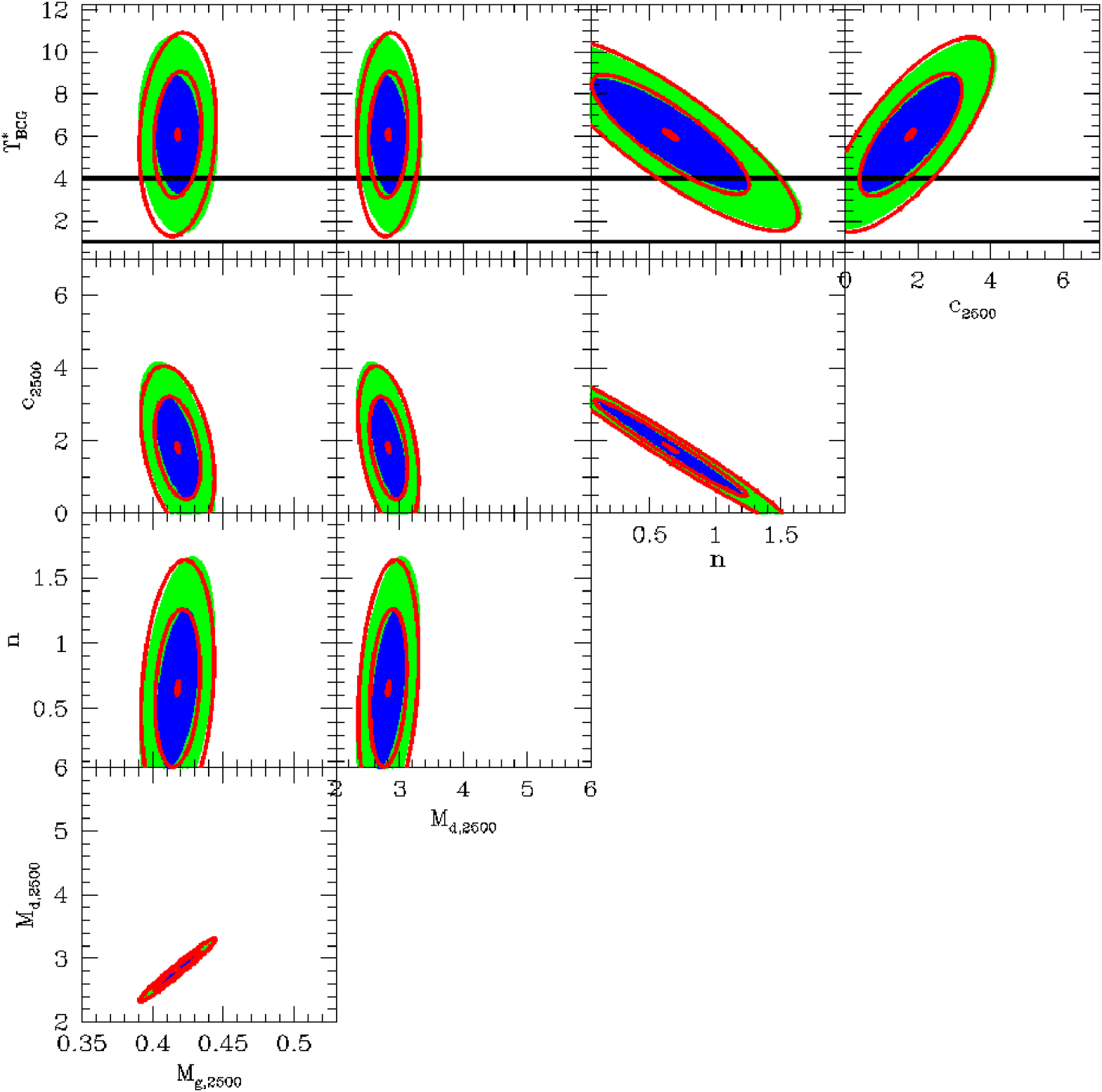}} &
\resizebox{\figreallyhalf}{!}{\includegraphics{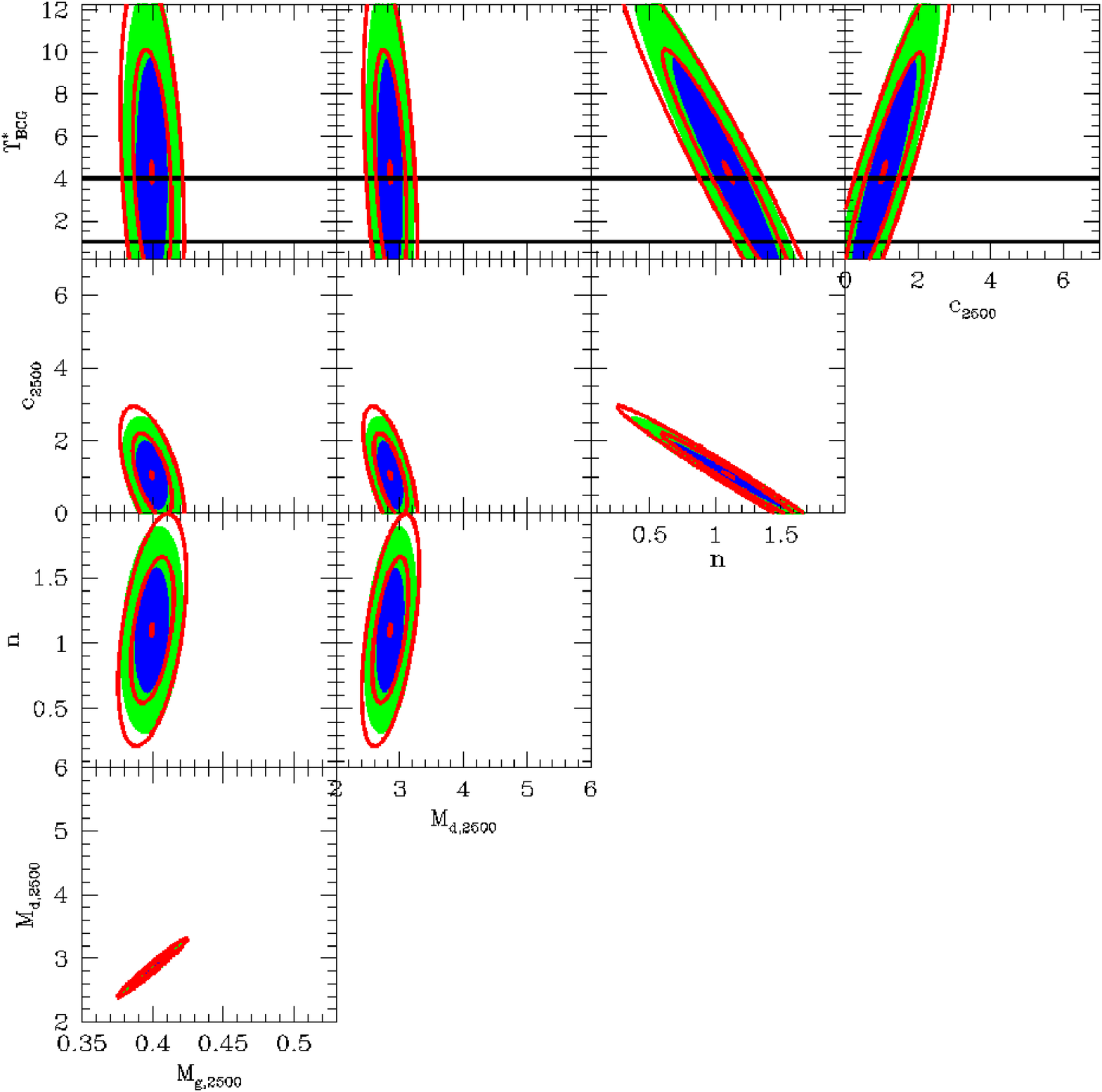}}
\end{tabular}
\figcaption{68\% and 95\% confidence contours on the dark mass within
$r_{2500}$, the gas mass within $r_{2500}$, the smallest of the three
$\beta$ parameters $\beta_1$, the concentration with respect to
$r_{500}$, the slope of the dark potential $n$, and the stellar
mass-to-light ratio of the BCG $\Upsilon^*_\mr{BCG}$. All masses are in units
of $10^{14} M_\odot$.  Clockwise from the top left are the Chandra
ACIS-S, XMM-Newton MOS1, XMM-Newton pn, and XMM-Newton MOS2
instruments. The filled contours show joint X-ray, SZ, and weak
lensing constraints. The unfilled contours show the X-ray constraints
alone. The thick lines show the allowed values (from theory) for
$\Upsilon^*_\mr{BCG}$.
\label{fig:confs}}

\end{figure*}
The breakdown for each instrument as well as the number of radial bins
appear in Table \ref{tbl:a478fit}. We show $\sim 25\%$ of the X-ray
spectra along with the best-fit spectral models in Figure
\ref{fig:somespectra}. We also show collapsed views of all the
spectral profiles in Figure \ref{fig:a478prof}.  The number of radial
bins exceeds those in previous mass measurement papers
\nocite{Pointecouteau04,Voigt06,Vikhlinin06}({Pointecouteau} {et~al.} 2004; {Voigt} \& {Fabian} 2006; {Vikhlinin} {et~al.} 2006) by a factor of $\sim 5$.
This allows us to constrain the dark mass and gas mass (or surface
brightness and temperature) profiles simultaneously, and ensures that
we do not miss any relevant details in the spatial distribution of the
spectra. The simultaneous fits to the SZ and weak lensing data
are shown in Figure \ref{fig:wlszdata}.

\begin{\dt}{lccccc}
\tablecaption{Best Fit Universal Models for  Abell 478 \label{tbl:a478fit}}
\tablewidth{0in}
\trot
\tablefontcommand
\tablehead{& \colhead{Chandra} & \colhead{EPIC MOS1} & 
\colhead{EPIC MOS2} & \colhead{EPIC pn}
& \colhead{Joint Constraints\tablenotemark}}
\startdata
\trot
X-ray $\chi^2$/DOF    &  5294/5413 ($q$=0.87)     &    4170/4084 ($q$=0.17)      &   6582/6326 ($q$=0.012)   &   7576/7462 ($q$=0.17) & 23264/23345  ($q$=0.63)   \\
Joint $\chi^2$/DOF   &  5659/5729 ($q$=0.74)     &    4526/4400 ($q$=0.090)     &  6936/6642 ($q$=0.006)   &  7929/7778 ($q$=0.11) & 23619/23661  ($q$=0.57)   \\
$M_{g,2500}$ & 0.403 $\pm$ 0.020 & 0.434 $\pm$ 0.013 & 0.414 $\pm$ 0.011 & 0.399 $\pm$ 0.009 & 0.425 $\pm$ 0.007 \\
$M_{d,2500}$ & 4.273 $\pm$ 0.753 & 3.254 $\pm$ 0.263 & 2.748 $\pm$ 0.199 & 2.839 $\pm$ 0.164 & 3.181 $\pm$ 0.135 \\
$n$          & 0.709 $\pm$ 0.436 & 1.035 $\pm$ 0.312 & 0.665 $\pm$ 0.346 & 1.096 $\pm$ 0.322 & 1.012 $\pm$ 0.207 \\
$c_{2500}$   & 2.027 $\pm$ 1.270 & 0.836 $\pm$ 0.528 & 1.966 $\pm$ 0.893 & 1.045 $\pm$ 0.633 & 1.234 $\pm$ 0.489 \\
$Z_0$        & 0.931 $\pm$ 0.153 & 1.960 $\pm$ 0.474 & 0.731 $\pm$ 0.228 & 1.440 $\pm$ 0.718 & 1.082 $\pm$ 0.416 \\
$Z_\infty$   & 0.095 $\pm$ 0.488 & 0.350 $\pm$ 0.144 & 0.159 $\pm$ 0.114 & 0.443 $\pm$ 0.069 & 0.336 $\pm$ 0.065 \\
$\Upsilon_\mr{BCG}$& $<3$        & 2.573 $\pm$ 2.217 & 5.487 $\pm$ 1.829 & 4.280 $\pm$ 3.251 & 1.669 $\pm$ 1.390 \\
\enddata
\tablecomments{Joint constraints on parameters for the
Universal fit to each dataset and to the joint data sets, including SZ
and weak lensing observations. We also show the goodness of fit $q$
for the X-ray observations alone. For a description of the parameters and their
units, see Table
\protect\ref{tbl:params} and \S\ref{sec:models}.  The confidence
intervals were obtained via the fit covariance matrix.}
\end{\dt}
The constraints from the fits appear in Figure \ref{fig:confs}, where
we show 68\% and 95\% the confidence intervals for some of the chief
parameters of interest. All plots assume a universal profile with free
inner slope $n$ (equation \ref{eq:universal}). We show the constraints
both with and without the inclusion of the weak lensing and SZ data.

The differences in the calibration among the various four instruments
are immediately apparent in the X-ray fits. For example, as is typical
of rich, hot clusters, the Chandra data yield a higher average
temperature \nocite{Kotov05,Vikhlinin05}({Kotov} \& {Vikhlinin} 2005; {Vikhlinin} {et~al.} 2005), resulting in a higher dark
mass measurement.  Also notable is the fact that the errors in the
Chandra-derived quantities are larger. Even though the Chandra ACIS-S
has the superior spatial resolution, XMM-Newton has the greater sky
area, so that the constraints on the shape of the dark profile, and
especially on the masses at $r_{2500}$, are tighter for the XMM-Newton
data. This is due to the fact that our outer extraction radius for
Chandra (450 kpc) is at about $0.6 r_{2500}$; quantities computed
outside this radius are extrapolations and therefore subject to
greater uncertainty.

It is also clear from Figure \ref{fig:confs} that the addition of the
lensing and SZ data can help constrain the dark matter parameters
significantly.  For example, the uncertainty in the dark matter
concentration as measured by Chandra is halved through the addition of
the lensing and SZ data. We explore the additional power afforded by
these types of observations in \S\ref{sec:extra} below. Most
encouragingly, there is no bias or disagreement whatsoever in the 
joint fit with the additional data sources---the same physical model
can account for the X-ray, SZ, and weak lensing observations at the
same time.

Chandra is the instrument for which $n$ is most affected by the
addition of the SZ and lensing data. The dark matter slope as measured
by Chandra alone ($n < 1$) is consistent with the XMM-Newton value
($1.1 \pm 0.3$).  When we add the SZ and lensing data to the Chandra
data, the slope ($n = 0.7 \pm 0.4$) is brought into closer agreement
with XMM-Newton, in that the low $n=0$ solution is disfavored.

\section{Discussion}
\label{sec:discussion}

\subsection{Comparison with Previous X-ray Studies}

For the first time, we are able to simultaneously constrain all the
parameters of physical interest in the mass fitting process, some of
which are listed in Table \ref{tbl:a478fit}.  JACO allows us to
calculate the joint probabilities of quantities such as the dark mass,
the gas metallicity, and the absorbing hydrogen column
density. Determination of the \emph{covariance} of such parameters is
a unique property of the method, and cannot be easily be duplicated
with previous techniques.

\begin{\dt}{lcccccccccc}
\tabletypesize{\footnotesize}
\tablewidth{0in}
\trot
\tablecaption{Comparison with Previous Work (X-ray Data Only) \label{tbl:compare}}
\tablehead{
\colhead{} & \colhead{JACO} & \colhead{S05}
& \colhead{VF06} & \colhead{JACO} & \colhead{P05} & \colhead{JACO} & \colhead{V06} & \colhead{VF06} & \colhead{JACO} \\
\colhead{Parameters\tn{a}} & \colhead{Chandra} & \colhead{Chandra} &
\colhead{Chandra} & \colhead{XMM} & \colhead{XMM} &
\colhead{Chandra} & \colhead {Chandra} & \colhead{Chandra} & \colhead{XMM}\\
\colhead{} & \colhead{All Data} & \colhead{All Data} &
\colhead{All Data} & \colhead{All Data} & \colhead{All Data} &
\colhead{Excised} & \colhead{Excised} & \colhead{Excised} & \colhead{Excised}}
\startdata
$n$\tn{b}              & $<1$           & $0.35 \pm 0.22$ & $0.49 \pm 0.45$       & $1.1 \pm 0.3$ & $\equiv 1$      & $<1.33$        & \nodata        &$1.1^{+0.2}_{-0.6}$& $1.7 \pm 0.3$ 
\\$r_{2500}$           & $0.68 \pm 0.07$  & $0.62 \pm 0.07$ & $0.76^{+0.67}_{-0.19}$& $0.59 \pm 0.02$ & \nodata       & $0.67 \pm 0.07$& \nodata        &  \nodata          & $0.60 \pm 0.03$  
\\$c_{2500}$           & $3.4 \pm 2.1$    &    \nodata      & \nodata               & $1.1 \pm 0.7$   & \nodata       & $<10$          & \nodata        &  \nodata          & $<0.70$  
\\$M_{\mr{tot},2500}$  & $4.9 \pm 0.9$    &    \nodata      & \nodata               & $3.3 \pm 0.2$   & \nodata       & $4.7 \pm 0.8$  & $4.2 \pm 0.3$  &  \nodata          & $3.4 \pm 0.3$  
\\$r_{500}$            & $1.5 \pm 0.2$    &    \nodata      & \nodata               & $1.4 \pm 0.1$   & \nodata       & $1.4 \pm 0.3$  & $1.4 \pm 0.1$  &  \nodata          & $1.5 \pm 0.5$  
\\$c_{500}$            & $7.4 \pm 4.3$    &    \nodata      & \nodata               & $2.4 \pm 1.3$   & \nodata       & $8.6 \pm 7.0$  & $3.8 \pm 0.3$  &  \nodata          & $<2.0$  
\\$M_{\mr{tot},500}$   & $9.9 \pm 2.6$    &    \nodata      & \nodata               & $8.2 \pm 1.0$   & $7.6  \pm 1.1$& $9.2 \pm 3.8$  & $7.8 \pm 1.4$  &  \nodata          & $11 \pm 6$ 
\\$r_{200}$            & $2.2 \pm 0.4$    & $2.2 \pm 0.1$   & $3.0^{+14}_{-1.0}$    & $2.1 \pm 0.2$   & $2.1 \pm 0.1$ & $2.1 \pm 0.3$  & \nodata        &$  \approx9.2$     & $2.6 \pm 0.6$  
\\$c_{200}$            & $11 \pm 6$       & $7 \pm 2$       & $2.9^{+2.0}_{-2.8}$   & $3.6 \pm 1.7$   & $4.2 \pm 0.4$ & $13 \pm 9 $    & \nodata        & \nodata           & $<1.2$  
\\$M_{\mr{tot},200}$   & $13 \pm 4$       & $13 \pm 3$      & \nodata               & $12 \pm 2$      & $11 \pm 2$    & $12 \pm 3$     & \nodata        & \nodata           & $21 \pm 9$
\enddata
\tablenotetext{a}{Distances are in units of Mpc and masses are in units of
$10^{14} M_\odot$. In all sources, measurements at $r_{500}$ and $r_{200}$ are based on extrapolation from
data within those radii. ``Excised'' fits exclude the central 30kpc.}
\tablenotetext{b}{Measured for the dark mass profile only in our work, 
and for the total mass profile in the other papers. All papers assume a universal profile (equation
\protect\ref{eq:universal}) with either free or fixed slope $n$; VF06 use a power law of slope $n$
in their excised fit.}
\tablecomments{Shown are 68\% confidence intervals on the structural
parameters. JACO: This work; S05: \nocite{Sanderson05}{Sanderson} {et~al.} (2005); 
V06: \nocite{Vikhlinin06}{Vikhlinin} {et~al.} (2006); VF06: \nocite{Voigt06}{Voigt} \& {Fabian} (2006); P05: \nocite{Pointecouteau05}{Pointecouteau} {et~al.} (2005).
``Chandra'' or ``XMM'' refers to the observatory used by each of
the cited works. ``All Data'' means that no portion of the cluster
emission was removed during analysis; ``Excised'' means that the central
$\approx 20-30$ kpc was removed before analysis.}
\end{\dt}

\begin{figure*}
\resizebox{\figfull}{!}{\includegraphics{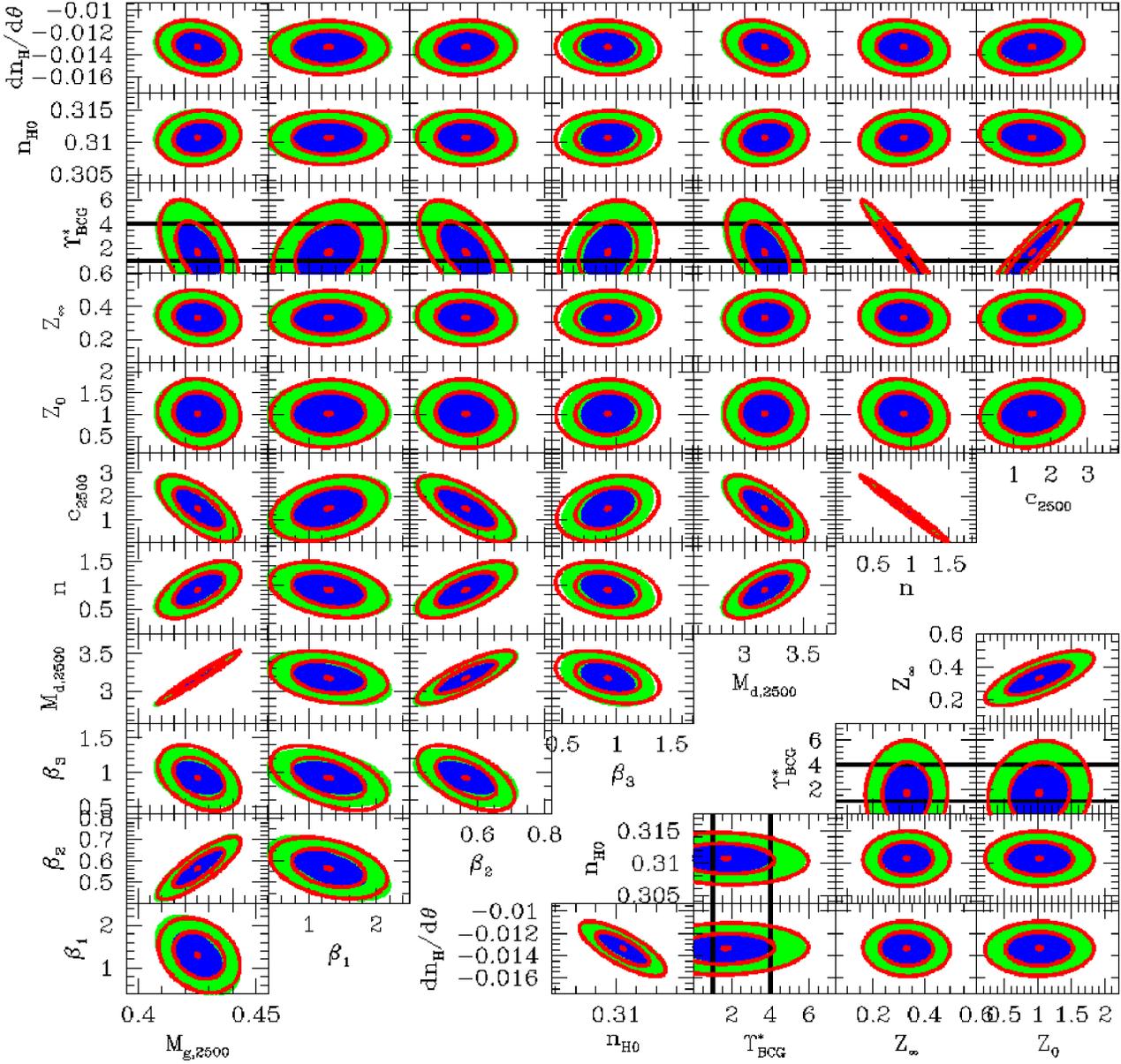}} \figcaption{68\%
and 95\% confidence contours for the combined JACO fit to all X-ray,
SZ, and weak lensing data. Shown are the dark mass within $r_{2500}$
(in units of $10^{14} M_\odot$), the gas mass within $r_{2500}$, the
three $\beta$ parameters $\beta_{1,2,3}$, the concentration with
respect to $r_{2500}$, the slope of the dark potential $n$, the
metallicity at the core ($Z_0$) and at large radii ($Z_\infty$), the
stellar mass-to-light ratio of the BCG $\Upsilon^*_\mr{BCG}$, the
central galactic absorption column $n_{H0}$ in units of $10^{22}$
cm$^{-2}$, and its gradient with respect to distance from the center
$\theta$ in arcminutes. The thick lines show the allowed values (from
theory) for the stellar mass-to-light ratio
$\Upsilon^*_\mr{BCG}$. \label{fig:final}}
\end{figure*}

\begin{figure}
\resizebox{\figszwl}{!}{\includegraphics{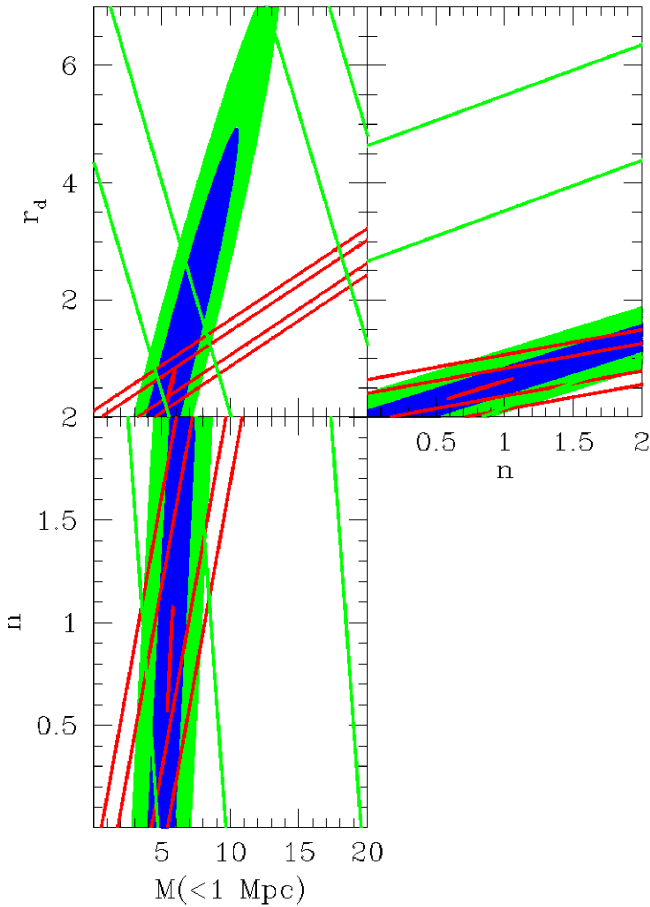}} \figcaption{Testing
the power of the SZ and weak lensing data without gas temperature
information. Shown are the dark mass within 1 Mpc in units of $10^{14}
M_\odot$, the dark radius from equation \protect\ref{eq:universal} in
units of Mpc, as well as the inner slope $n$.  The X-ray gas mass is
taken as a fixed prior, but X-ray temperature (spectral) information
is not used. Solid contours show SZ+weak lensing constraints; dark
unfilled contours show the lensing constraints alone, and the light
unfilled contours show the SZ constraints alone. \label{fig:szwl}}
\end{figure}
\begin{figure}
\resizebox{\fighalf}{!}{\includegraphics{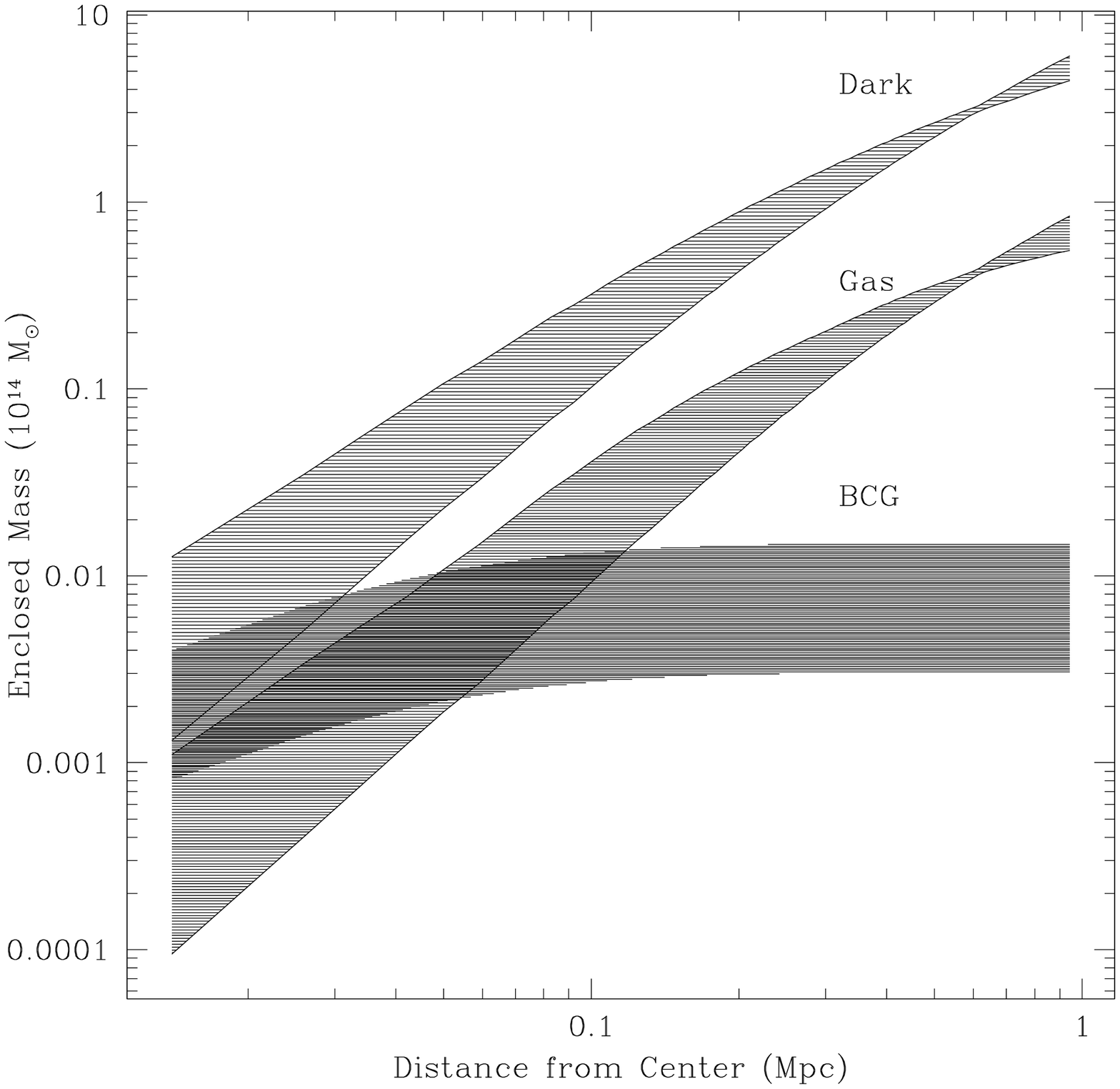}}
\figcaption{Total
enclosed mass for the dark, gaseous, and stellar BCG components of
Abell 478, in units of $10^{14} M_\odot$. Shown are the 68\%
confidence intervals for fit to all data (Chandra + XMM-Newton + SZ +
weak lensing). The stellar mass constraint includes the assumption
that the stellar mass-to-light ratio $\Upsilon^*_\mr{BCG} >
1$. \label{fig:massbudget}}
\end{figure}

In Table \ref{tbl:compare} we compare the chief structural properties
of the best-fit JACO model with previous studies of Abell 478.  In
general, the JACO analysis of the Chandra and XMM-Newton data is in
excellent agreement with previous works. In comparing the JACO Chandra
results with \nocite{Sanderson05}{Sanderson} {et~al.} (2005) and \nocite{Vikhlinin06}{Vikhlinin} {et~al.} (2006), we note that
the uncertainties in the measured JACO parameters are larger.  The
explanation for this is that JACO allows \emph{all} physical
quantities to vary simultaneously during the fitting process. Other
fitting techniques conduct a multi-stage fit: first they fix the
surface brightness profile at the best-fit multiple $\beta$ model, and
then, using this fixed profile, they fit a temperature profile to the
reduced spectra. This results in an underestimate of the true
uncertainties in $n$ and the mass, which are better reflected in the
JACO measurements.

We note that the JACO results isolate the dark matter from the gas and
stellar profiles. In other studies, the measured $n$ reflects the
total gravitational potential.  The similarity of the JACO $n$ values
compared with the globally measured $n$ in the literature yields the
useful conclusion that the stellar and gas profiles do not
significantly affect the measured dark matter slopes, at least for
Abell 478.

\subsection{Contamination by the Central Source}

A key problem with the mass measurement within the inner 20-30 kpc of
Abell 478 (and other similar cool core clusters) is AGN activity. The
symmetric bubbles discussed by \nocite{Sanderson05}{Sanderson} {et~al.} (2005) are clearly visible
in Figure \ref{fig:hst}.  All X-ray mass measurement techniques rely
on the assumption of hydrostatic equilibrium. However, in most relaxed
clusters there is clear, complex interaction between the central
AGN/radio source and the intervening gas
\nocite{Blanton03b,Clarke04,McNamara05}(e.g. {Blanton} {et~al.} 2003; {Clarke}, {Blanton}, \& {Sarazin} 2004; {McNamara} {et~al.} 2005).

In the discussion thus far, we have conducted a hydrodstatic analysis
including this central region. There is at this point no consensus on
whether hydrostatic analysis of nonequilibrium gas yields correct
results.  One common approach \nocite{Voigt06,Vikhlinin06}({Voigt} \& {Fabian} 2006; {Vikhlinin} {et~al.} 2006) is to excise
the inner region from analysis. Note that excising the X-ray emission
around the BCG does not mean setting the mass within that region equal
to zero---it merely means that the shape of the mass distribution
inside that region is not constrained by the X-ray model. For
XMM-Newton, excising only the central region is problematic: the large
size of the PSF and the steepness of the surface brightness profile
ensure that the central spectrum contributes as far out as $1\arcmin$
from the center \nocite{MarkevitchPSF}({Markevitch} 2002).

To examine the effects of removing the emission from the disturbed
region, we repeat our entire X-ray analysis, excising the central
20\arcsec\ (32 kpc) for Chandra, and the central 1\arcmin\ (100 kpc) for
the XMM-Newton data. The results appear in Table \ref{tbl:compare}. We
find that removing the central portion causes us to measure a steeper
value for $n$ in both the XMM-Newton and the Chandra data. 
Outside 100 kpc, the XMM-Newton dark matter profile is consistent with
a single power law (the concentration is consistent with 0 in all
cases). The Chandra dark profile becomes more fully consistent with an
$n=1$ Universal profile, as first pointed out by \nocite{Voigt06}{Voigt} \& {Fabian} (2006).

Given the quality of the Abell 478 observations, these results show
that strong constraints on the shape of the dark profile depend
greatly on the data within $\approx 20-30$ kpc of the center.
Unfortunately, we do not yet know whether the hydrostatic model
considered here gives correct results when applied to regions where
AGN heating is important. However, at least for Abell 478, the mass
profiles measured for the full data set are consistent with those
measured with the central 30 kpc excluded (the latter have larger
errors).  Thus, treating the central region as hydrostatic does not
appear to bias the final measured parameters.

\subsection{Combined Constraints from All Data Sources}

We now consider the question of conducting a grand total fit to all
four instruments. This is a problematic issue, because of the known
disagreement in the temperature profiles of this cluster. In
combination, the spectra from the four X-ray instruments, together
with the lensing and SZ observations, contain \fseall\ data points
(including over a million X-ray photons). To account for the
differences in the cross-calibration, the simultaneous fit to all four
instruments includes a 4\% systematic error added in quadrature to the
statistical error.  Without the addition of a systematic component no
model can simultaneously provide a good fit to both the Chandra and
XMM-Newton data. The magnitude of the systematic error was chosen by
comparing count ratios in published cross-calibration
studies \nocite{XMMcrosscal}({Stuhlinger} {et~al.} 2006). 

Figure \ref{fig:final} shows the combination of all four instruments
with the X-ray, SZ, and weak lensing data for the Universal
profile. The total number of degrees of freedom are \fsedof. We find
that the $n\sim1$ universal profile provides the best overall fit,
>with $\chi^2 = 23619$. The $n\sim1$ gamma-profile \nocite{Hernquist90}(essentially
a {Hernquist} 1990, profile) provides the next best match, with $\Delta
\chi^2=15$ with respect to the universal profile. The constant baryon
fraction model and the single power law models provide worse matches
to the data, with $\Delta\chi^2 = 37$ and $\Delta\chi^2 =187$,
respectively. Using the likelihood ratio test, and noting that a
single power law is obtained from the universal profile by setting
$r_d=0$, we estimate that the single power law model is disfavored
with false-alarm probability $p < 10^{-6}$.

\subsection{The Leverage of SZ and Weak Lensing Data}
\label{sec:extra}
We showed in \S\ref{sec:results} that the addition of SZ and weak
lensing data results in improved constraints on the shape parameters
of the gas and dark matter density (at least for the Chandra data
taken alone). Here we examine the reasons for this improvement in
further detail. As an exercise, we fix the gas mass distribution at
the profile derived from the fit to the combined X-ray data.  We allow
only the dark and stellar mass profiles to vary. Thus we take
advantage of the fact that the gas mass profile is a well measured and
well agreed-upon quantity for most clusters. For this exercise we
ignore the spectral information, where the Chandra and XMM-Newton
disagreement is more serious.

The results appear in Figure \ref{fig:szwl}. These figures show that
the SZ and weak Lensing data are complementary, as suggested in
previous theoretical studies \nocite{Zaroubi01,Dore01}({Zaroubi} {et~al.} 2001; {Dor{\'e}} {et~al.} 2001). While neither
can by itself strongly constrain the slope and concentration of the
dark matter profile, together they offer nearly orthogonal constraints
on these parameters.

Comparing Figure \ref{fig:szwl} to Figure \ref{fig:confs}, it is clear
that whenever the quantity of X-ray data overwhelms the SZ and weak
lensing data, the statistical weight of the latter two is small; this
is why the high quality EPIC pn and MOS2 results are nearly unchanged
by the addition of the SZ and weak lensing information. However, when
the spatial resolution and the number of X-ray photons is small---as
is surely the case for most higher redshift clusters---the SZ and weak
lensing information contribute substantially to the constraints on the
dark matter profile.

\subsection{Dark Matter and the Role of the BCG}

The stellar mass of the BCG is an important consideration for modeling
the inner regions of galaxy clusters \nocite{Lewis03,Mamon05}({Lewis} {et~al.} 2003; {Mamon} \& {{\L}okas} 2005). In
Figure \ref{fig:final}, the correlation between the dark matter slope
$n$ and the stellar mass-to-light ratio $\Upsilon^*_\mr{BCG}$ is strong.  The
more massive the BCG, the less room there is for a central dark matter
cusp. On the other hand, for a given BCG luminosity, there is a
minimum value of $\Upsilon^*_\mr{BCG}$ regardless of the stellar initial mass
function (IMF). The star light in effect yields an upper limit on the
steepness of the dark matter profile.

A galaxy dominated by a $> 3$ Gyr old stellar population ought
to have $\Upsilon^*_\mr{BCG} \approx 1-4$ in the HST F606W band
\nocite{Maraston98}({Maraston} 1998). These limit are shown in Figures \ref{fig:confs}
and \ref{fig:final}. Adopting these limits yields further constraints
on the dark matter slope $n$ (towards lower values) and the
concentration (towards higher values). For example, in Figure
\ref{fig:final}, values of $n > 1.1$ would be ruled out by the allowed
$\Upsilon^*_\mr{BCG}=1$ limit at the 99\% confidence level. The ruled-out
region changes to $n > 1.3$ if when we use the excised data only. It
is worth noting that while the \nocite{Maraston98}{Maraston} (1998) population synthesis
models have different stellar mass-to-light ratios depending on the
exact mixture of stellar populations, only the upper limit
($\Upsilon^*_\mr{BCG} = 4$) is sensitive to the shape of the initial mass
function; the lower limit, $\Upsilon^*_\mr{BCG}=1$, is firm and independent of
the IMF for a given age.

Figure \ref{fig:massbudget} shows the mass fraction in stars for Abell
478 as a function of distance from the center.  As a contributor to
the equation of hydrostatic equilibrium, the stars in the BCG are as
important or more important than the gas within 80 kpc. The stars also
contribute 30\% of the total mass within 20 kpc of the center.

\section{Conclusion}
\label{sec:conclusion}

We have developed a new method for measuring the dark matter profile
of a cluster directly from X-ray, lensing, and SZ data.

\begin{itemize}
\item JACO works directly in the data plane, and generates the
observables (X-ray spectra, weak lensing shear profiles, or
Sunyaev-Zel'dovich decrement maps) from candidate models. It therefore
allows simultaneous constraints on a cluster's dark matter profile
using multiwavelength data. It allows joint constraints on all
physical parameters of interest.

\item JACO explicitly separates the dark, gas, and stellar mass
profiles. This allows extraction of the shape of the dark profile
independently of the rest of the cluster.

\item As long as the gas mass profile is well known, the SZ and weak
lensing data together can provide orthogonal constraints on the shape
parameters of the dark profile.

\item We present new CFHT weak lensing measurements for the
well-studied rich cluster Abell 478, and provide an improved reduction
of existing CBI Sunyaev-Zel'dovich data. We analyze these data in
conjunction with existing Chandra and XMM-Newton observations of the
cluster. We find excellent agreement among all data sets when they are
fit simultaneously with JACO models. The weak lensing and SZ
observations improve the constraints on the shape of the dark matter
distribution from the Chandra data.

\item The Chandra and XMM-Newton data for Abell 478 yield similar
slopes for the inner dark matter profile: $n < 1$ and $n = 1.1 \pm
0.3$ at the 68\% confidence level, respectively. The Chandra
constraints become more fully consistent with the $n=1$ NFW profile
when we excise the morphologically disturbed central 30 kpc. A similar
result was noted by \nocite{Voigt06}{Voigt} \& {Fabian} (2006). 

\item At intermediate and higher redshifts, where the quality of the
X-ray data rapidly diminishes, the SZ and weak lensing data become
increasingly important for characterizing the properties of dark
matter. In these regimes JACO will be a powerful tool for improving
constraints on the shape of the cluster potential.

\item JACO as described in this paper will soon be available for
public download at \url{http://jacocluster.sourceforge.net}. 

\end{itemize}

In future papers in this series, we will test JACO against
gasdynamical N-body simulations, generalize the method to the
axisymmetric case, and use JACO to analyze multiwavelength data for a
large cluster sample.

\acknowledgments

We are grateful for comments from the referee, who led us to improve
the paper. We thank Alexey Vikhlinin, Trevor Ponman, George Lake, and
Dick Bond for insightful discussions. AM acknowledges partial support
from a Chandra Postdoctoral Fellowship issued by the Chandra X-ray
Observatory Center, which is operated by the Smithsonian Astrophysical
Observatory for and on behalf of NASA under contract NAS 8-39073. AB
and HH acknowledge NSERC Discovery Grants.  HH acknowledges support
from the Canadian Institute for Advanced Research. We acknolwedge use
of equipment acquired using grants from the Canada Foundation for
Innovation and the British Columbia Knowledge and Development
Fund. This research used the facilities of the Canadian Astronomy Data
Centre operated by the National Research Council of Canada with the
support of the Canadian Space Agency.  This work is based on
observations obtained at the Canada-France-Hawaii Telescope (CFHT)
which is operated by the National Research Council of Canada, the
Institut National des Sciences de l'Univers of the Centre National de
la Recherche Scientifique of France, and the University of Hawaii.
This work makes substantial use of version 1.7 of the GNU Scientific
Library, an excellent library of numerical routines in C distributed
under the GNU General Public License.

\appendix 
\section{Fast evaluation of broken power law integrals}
\label{sec:incompletebeta} 

To calculate the integrated mass of the profiles considered in 
\S\ref{sec:models}, we often need to calculate integrals of the
type
\begin{equation}
M(r) = \int_0^r 4 \pi r^2 \rho(r/r_t) dr = 4 \pi r_t^3 \rho_0 \int_0^{r/r_t} x^{2-n} (1+x^p)^{\frac{n-q}{p}} dx
\end{equation}
where $r_t$ is the characteristic radius of the density $\rho$, and
$n$ and $q$ are the slopes at $r=0$ and $r=\infty$,
respectively. Rather than resorting to hypergeometric functions to
evaluate the integral, we can express it in terms of the incomplete
beta function $B_y(a,b)$ for which very fast numerical routines exist:
\begin{equation}
M(r)  =  4 \pi r_t^3 \rho_0 B_y \left(\frac{3-n}{p},\frac{q-3}{p} \right); 
y  \equiv  \frac{r^p}{r_t^p + r^p}
\end{equation}
In the common definition of the incomplete beta function, $a$ and $b$
must be positive definite. However, $q-3$ could well be negative
(e.g., for a $\beta$-model).  The following recurrence relations allow
us to transform the $2 < q < 3$ cases into incomplete beta functions with
positive definite arguments:
\begin{eqnarray}
\label{eq:transform}
B_y(a,b) & = & B(b,a)-B_{1-y}(b,a) \\
B_y(a+1,b) & = & \frac{1}{a+b} \left[ a B_y(a,b)- (1-y)^b y^a \right]
\end{eqnarray}
where $B(a,b) = B_0(a,b)$ is the complete beta function. For $y <
(a+1)/(a+b+2)$, $B_y(a,b)$ is evaluated using rapidly converging
continued fractions. For $y > (a+1)/(a+b+2)$, equation
(\ref{eq:transform}) is used to transform the problem back into a
regime where the continued fractions converge quickly.

For the specific case that $b = 0$ (e.g. NFW profiles), the
transformations cannot be used. However, the continued fraction method
yields results accurate to better than $10^{-6}$ for all interesting
values of $y$ so long as $a$ is limited to physically plausible values
(i.e. is limited to 1 to 3, corresponding $n = $ 0 to 2).

\section{Calculation of the PSF Coefficients}
\label{sec:psfintegral}

Here we discuss the approximations used for calculating the scattering
of light from an annulus with inner and outer radii ($R_{j-1}$,$R_j$)
into an annulus with inner and outer radii ($R_{i-1}$,$R_{i}$). The
general expression for $C\nu(x,y)$, the counts observed using a
detector with a monochromatic point spread function
$p_\nu(x,y,x\p,y\p)$ observing a monochromatic source with flux
distribution $I_\nu(x,y)$ is
\begin{equation}
C_\nu(x,y) = \int \int I_\nu(x\p,y\p) p_\nu(x,y,x\p,y\p) dx\p dy\p
\end{equation}
Approximating the source and the PSF as azimuthally symmetric, the
above expression becomes
\begin{equation}
C_\nu(R) = \int_0^{2 \pi} \int_{0}^{\infty} R\p I_\nu(R\p) p_\nu(R,R^2+R^{\prime 2}- 2 R
R\p \cos{\phi} ) dR\p d\phi,
\end{equation}
where $R^2+R^{\prime 2}- 2 R R\p \cos{\phi}$ is the square of the
distance between $(x,y)$ and $(x\p,y\p)$ in polar
coordinates. Splitting up the source flux $I(R\p)$ into $N$
homogeneous annuli, we have
\begin{equation}
C_\nu(R) = \sum_{j=1}^N I_\nu(R_{j-1},R_j) \int_0^{2 \pi}
\int_{R_{j-1}}^{R_j} R\p I(R\p) p_\nu(R,R^2+R^{\prime 2}- 2 R
R\p \cos{\phi} ) dR\p d\phi
\end{equation}
where $R \equiv (R_{i-1}+R_i)/2$. Each set of $N$ double integrals
yields the PSF correction coefficients at energy h$\nu$. We evaluate
the double integral for four photon energies ($h \nu =
0.37,0.75,1.5,3.6$ keV), and interpolate to obtain the correction
at arbitrary energy. 

\section{Technical Details of the Reduction and Analysis}
\label{sec:reduction}

JACO v1.0 includes scripts which prepare the raw data from the Chandra
and XMM-Newton archives for analysis. These preprocessing scripts
yield a uniformly reduced set of spectra for an entire cluster with
minimal human interaction. The use of these scripts is not required
for undertaking the JACO analysis. Observers may undertake any
reduction procedure in preparation for JACO, subject to the following
requirements on the final output spectra:
\begin{enumerate}
\item Each spectrum must be extracted from a circular annulus
surrounding the cluster center. Future versions of JACO will also
handle elliptical annuli.

\item The spectrum keyword BACKSCAL must be set to the net source area
in square arcminutes (i.e. excluding the area of any excised regions
intersecting the annulus).

\item The redistribution matrix file (RMF) and ancilliary response
file (ARF) must have the same energy binning for all spectra. The
largest bin size that preserves relevant calibration features should
be used.

\end{enumerate}

\subsection{Pipeline reprocessing} 

To begin, the JACO preprocessing scripts reapply the latest version of
the standard pipeline analysis to the raw data. For Chandra, this has
the effect of correcting for the spatial gradients in the filter
contamination. For ACIS front-illuminated chips with the appropriate
detector temperature, we include the charge transfer inefficiency
correction (CTI) \nocite{Townsley00}({Townsley} {et~al.} 2000). For XMM-Newton, we rerun the
standard pipeline to include the necessary files for correcting the
EPIC pn spectra for the out-of-time (OOT) events.

\subsection{Point source removal} 

Because we are interested only in the diffuse X-ray emitting medium,
we remove contaminating point sources from the data. The CIAO wavelet
source detection algorithm yields source lists for both the Chandra
and the XMM-Newton total band images.  Visual inspection and
correction of the source lists is necessary to remove spurious
detections along chip borders. For each detected point source, we
remove an elliptical region with major axis equal to three times the
PSF width from photon event files. The source detection program
specifies the orientation and axis ratio of the ellipses.
\subsection{Flaring background} 

Flares in the X-ray background can contaminate the spectra and must
therefore be screened using the field total light curve at energies
$\gtrsim 10$ keV.  In many XMM-Newton observations, a substantial
fraction of the telescope live time is affected by flares. To screen
the XMM-Newton data, we apply 2$\sigma$ clipping of the light curve
binned in 100s intervals. Then we apply the technique of
\nocite{Nevalainen05}{Nevalainen}, {Markevitch}, \&  {Lumb} (2005). They find that small flares significantly affect
the $\sigma$-clipped XMM-Newton data, and that these flares are best
filtered by accepting only periods with count rates less than 120\% of
the mean $\sigma$-clipped count rate. For Chandra data, we follow the
flare rejection method provided as a standard contributed background
analysis package \nocite{MaximCookbook}({Markevitch} 2005).

\subsection{Quiescent particle background}

Two types of background affect Chandra and XMM-Newton observations of
diffuse sources. The non X-ray particle background is dominated by the
interaction of charged and non-charged particles with the CCD. In both
observatories, the particle background comprises a continuum as well
as spectral line features whose location depends on the particular
instrument. Being non X-ray in origin, the particle background is not
vignetted---i.e., it is not affected by the variations in the
effective area across the CCD. To subtract this background from the
spectra, we use the latest public versions of the blank sky
observations for XMM-Newton \nocite{Read03}({Read} \& {Ponman} 2003) and Chandra
\nocite{MaximCookbook}({Markevitch} 2005). The subtraction is made possible by the fact
that after flare rejection, there is little change in the spectral
shape of the particle background with position or time; only the
normalization varies appreciably.

To carry out the subtraction, we subject the XMM-Newton blank sky
spectra to same flare rejection algorithm as the cluster spectra (the
Chandra blank sky are already calibrated in the same manner as the
observations). We normalize the spectra of the blank sky fields to
match the spectra of the cluster fields observed in the 10-12 keV
(XMM-Newton) and 8-10 keV (Chandra) energy range (where the effective
area for X-ray photons is low). 
For
the purposes of calculating the renormalization factor, we exclude
emission from the central 100 kpc of the cluster.  
For each extracted cluster spectrum, we use the blank sky fields to
extract a matching particle background spectrum. 

\subsection{Residual X-ray Background} 

Once point sources are removed, the remaining diffuse emission
consists of two components: unresolved extragalactic AGN (with index
$\sim 1.4$ power law spectra) and unabsorbed, thermal $\sim 0.2$
keV background with approximately solar abundance
\nocite{Markevitch03b}({Markevitch} {et~al.} 2003). 
Unlike the particle background, the soft X-ray background
varies with position on the sky, so that it cannot be corrected using
blank sky fields. 

One option for dealing with the residual background is to ignore it
completely by cutting all emission below 1 keV. However, this also
removes a significant portion of the cluster signal, including
potentially important low-energy line emission. Another option is to
measure the spectrum in a source-free region and subtract it from the
cluster \nocite{Read03}({Read} \& {Ponman} 2003). However clusters of galaxies often fill the
entire field of view of the X-ray detector; it is difficult to find a
truly source-free region.

Our approach is to model the residual background as part of the
fitting process \nocite{Mahdavi05}(e.g. {Mahdavi} {et~al.} 2005). This is the only realistic
option when the cluster fills the entire field of view.  For each
instrument, we add a different unabsorbed thermal plasma with free
temperature and abundance and arbitrary (negative or positive)
normalization. To account for the diffuse extragalactic background, we
add an absorbed power law with slope fixed to 1.4 and similarly free
normalization. We require the temperature and metal abundance to be
the same for all instruments. However, different normalizations are
required for each detector because the instrumental background from
the blank sky fields is different for each instrument. When the
renormalized instrumental background is subtracted, the net (positive
or negative) cosmic background in each field will be different, even
though the true cosmic background is the same.

\subsection{Spectrum Extraction}

At present the JACO method includes only spherically symmetric
analysis; we therefore extract spectra in concentric annuli around the
cluster center. The cluster center is determined using the peak of the
Chandra X-ray emission, and verified using Hubble Space Telescope 
archival data.

JACO requires a minimum annulus width of 5\arcsec. At these widths,
the PSF correction (\S\ref{sec:psf}) becomes important even for
Chandra data. The widths are by default set to be the same for all the
instruments considered.  The JACO user can set the minimum number of
background-subtracted counts required per annulus. The background
measured at this stage (and this stage only) is roughly estimated from
the region outside the largest circular annulus that will completely
fit in the instruments' field of view. The spectra are grouped by
default at 40 events per bin, but the user can modify this.

To model the spectra, an accurate knowledge of the total solid angle
subtended by each annulus is necessary. Because an annulus can
intersect chip gaps, bad pixels, and excised sources, the calculation
of this area is not trivial. In particular, the XMM-Newton SAS
\verb=backscale= procedure cannot calculate accurate areas for the
annuli with the smallest (5\arcsec) width. For this reason, JACO
creates its own high resolution bad pixel map for each of the three
EPIC instruments, and uses it to determine the correct solid angles.

We use the appropriate CIAO and SAS tasks to create response matrices
for the spectra. For Chandra, we use \verb=mkwarf/mkacisrmf=; for
XMM-Newton, we use \verb=arfgen/rmgfen=. The response functions are
emission-weighted through the use of weight maps. \nocite{Arnaud01}{Arnaud} {et~al.} (2001)
argue that this technique is imperfect, because the spatial variation
of the uncalibrated source can yield biases in the X-ray
temperatures. Their solution, \verb=evigweight=, is an alternative
technique for the reduction of XMM-Newton spectra. In
\verb=evigweight= the individual photon events are weighted by the
inverse of the effective area at the photons' recorded positions, and
only one central ARF is created for the entire cluster. Because CIAO
does not offer a similar procedure for Chandra spectra, we have
compared XMM-Newton spectra derived using the standard method with
spectra derived using \verb=evigweight=.  We find that for the fine
binning used by JACO, the difference between the standard and
\verb=evigweight= spectra are negligible as long as the detector map
is extracted for energies where the raw counts spectrum is closest to
constant, specifically the channels corresponding to the 0.8-1.4 keV
photon energy range.

\subsection{Covariant Error Analysis: the Hrothgar Parallel Minimizer}

The command-line (as opposed to the Sherpa-based) frontend to JACO
allows the global $\chi^2$ to be minimized within a clustered
computing environment. In essence, Hrothgar runs the
Levenberg-Marquard (LM) minimization algorithm \nocite{Press92}({Press} {et~al.} 1992) as
implemented by the GNU Scientific Library (GSL;
\url{http://www.gnu.org/software/gsl/}).  The LM algorithm requires
knowledge of the Jacobian of the function to be minimized, which is
not analytic for X-ray spectra due to the highly nonlinear nature of
the cooling function. Hrothgar calculates the Jacobian numerically
using the methods described in \nocite{Press92}{Press} {et~al.} (1992), with error control as
implemented by the GSL library. Because calculation of the Jacobian
matrix $J$ during the minimization phase requires $2m$ evaluations of
the JACO model for $m$ free parameters, each function evaluation can
be assigned to a cluster node, resulting in a speed increase directly
proportional to the number of CPUs, up to $2m$ CPUs.

We carry out the error analysis by calculating the covariance matrix
$C = (J J^T)^{-1}$. When computing $J$ numerically for this purpose,
we use both the 3-point and the 5-point numerical differentiation
rules. The difference between the 3- and the 5-point rules yields an
estimate of the error, which we minimize as a function of the
step size. The covariance matrix obtained in this manner records the
interdependence of all the physical parameters in the JACO
analysis. The confidence contours are plotted by the methods described
in \nocite{Press92}{Press} {et~al.} (1992).

Hrothgar will soon be freely downloadable at
\url{http://hrothgar.sourceforge.net}.

\ifthenelse{\isundefined{\usemyrefs}}{
\section*{References}


}{ 
\bibliography{} }

\end{document}